\documentclass{aa}  

\usepackage{graphicx}
\usepackage{natbib}
\begin{document}

   \title{Particle acceleration and transport in reconnecting twisted loops in a stratified atmosphere}

   \author{M. Gordovskyy \inst{1}\fnmsep\thanks{\email{mykola.gordovskyy@manchester.ac.uk}}, 
	  P.K. Browning \inst{1}, E.P. Kontar \inst{2}
          \and
          N.H. Bian\inst{2}}

   \institute{Jodrell Bank Centre for Astrophysics, University of Manchester, Manchester M13 9PL, UK\\
         \and
             School of Physics and Astronomy, Univesity of Glasgow, Glasgow G12 8QQ, UK\\
             }

   \date{Received ; accepted }

\authorrunning{Gordovskyy et al.}
\titlerunning{Particle acceleration and transport in twisted loops}

% \abstract{}{}{}{}{} 
% 5 {} token are mandatory
 
  \abstract
  % context heading (optional)
  % {} leave it empty if necessary  
   {Twisted coronal loops should be ubiquitous in the solar corona. Twisted magnetic fields contain 
	excess magnetic energy, which can be released during magnetic reconnection, causing solar flares.}
  % aims heading (mandatory)
   {The aim of this work is to investigate magnetic reconnection, and particle acceleration and transport in 
	kink-unstable twisted coronal loops, with a focus on the effects of resistivity, loop geometry 
	and atmospheric stratification. Another aim is to perform forward-modelling of bremsstrahlung emission and determine the 
   structure of hard X-ray sources.}
  % methods heading (mandatory)
   {We use a combination of magnetohydrodynamic (MHD) and test-particle methods. First, the evolution of the kinking coronal loop 
	is considered using resistive MHD model, incorporating atmospheric stratification and loop curvature. 
   Then, the obtained electric and magnetic fields and density 
	distributions are used to calculate electron and proton trajectories using a guiding-centre approximation, taking
   into account Coulomb collisions.}
  % results heading (mandatory)
   {It is shown that electric fields in twisted coronal loops can effectively accelerate protons and 
	electrons to energies up to 10 MeV. High-energy particles have hard, nearly power-law energy 
	spectra. The volume occupied by high-energy particles demonstrates radial expansion, which results in
	the expansion of the visible hard X-ray loop and a gradual increase in hard X-ray footpoint area. Synthesised
   hard X-ray emission reveals strong footpoint sources and the extended coronal source, whose intensity
   strongly depends on the coronal loop density.}
  % conclusions heading (optional), leave it empty if necessary 
   {}


   \maketitle
%
%________________________________________________________________

\section{Introduction}\label{intro}

Twisted magnetic fields should be ubiquitous in the solar corona. Various types of photospheric motions, such as shear and
rotation \cite[often observed in sunspots, see for example][]{broe03} add helicity to the coronal field structure. At the same time,
the new emerging magnetic flux requires a non-zero helicity in order to form a loop, instead of a diffuse magnetic field 
arcade \cite[see simulations by][and discussion therein]{arce10}. Twisted coronal loops are often directly observed in extreme 
ultra-violet (EUV) \citep[see e.g.][]{raou09,srie10} and found in reconstructed coronal magnetic fields 
\citep[e.g.][]{ream04, male11}.

The twisted magnetic field contains excess magnetic energy which can be released with the field relaxing towards a potential 
configuration. For this reason twisted fluxtubes in the corona are often associated with solar flares and other active 
events. A number of observational works reveal the presence of a twisted magnetic field in pre-flare active regions; 
some of them directly associate the twisted magnetic ropes with the energy release \citep[see e.g.][]{srie10}. Furthermore, 
there seems to be a strong link between kink and torus instabilities in the twisted coronal ropes, and the 
ejection of plasmoid and a subsequent coronal mass ejection (CME), which is confirmed by observational evidence 
and simulations \citep[see][]{amae00,tokl05,gibe06,fan10}. However, reconnection 
in twisted magnetic loops in the corona could also explain events with no plasmoid ejection or CME. \citet{asce09} showed that a
substantial number of small flares occur in a simple single loop configuration, also known as self-contained flares. It is 
suggested that magnetic reconnection in twisted coronal loops in an active region with no open field could explain these phenomena.

There are several studies concerning magnetohydrodynamic (MHD) properties of twisted coronal loops. Kliem, Toeroek, and colleagues 
investigated kink and torus instabilities in twisted loops in zero-beta plasma \citep{tokl05,klie00,klie10}. The model was later 
applied to study 
the ejective eruptions and CME. \citet{brva03} and \citet{broe08} considered stability of twisted 
magnetic loops using a semi-analytical approach, determining criteria for ideal kink instability and subsequent relaxation
during a nonlinear phase of the instability. These results have been 
recently generalised by \citet{bare13}. These criteria has been used by \citet{broe08}, \citet{hooe09}, and later by \citet{gobr11} 
to perform numerical simulations of magnetic reconnection in a kink unstable magnetic fluxtube. All these studies consider
instability and magnetic reconnection in a fluxtube, which is cylindrically symmetric.  

\begin{figure*}
\centerline{\includegraphics[width=1.0\textwidth]{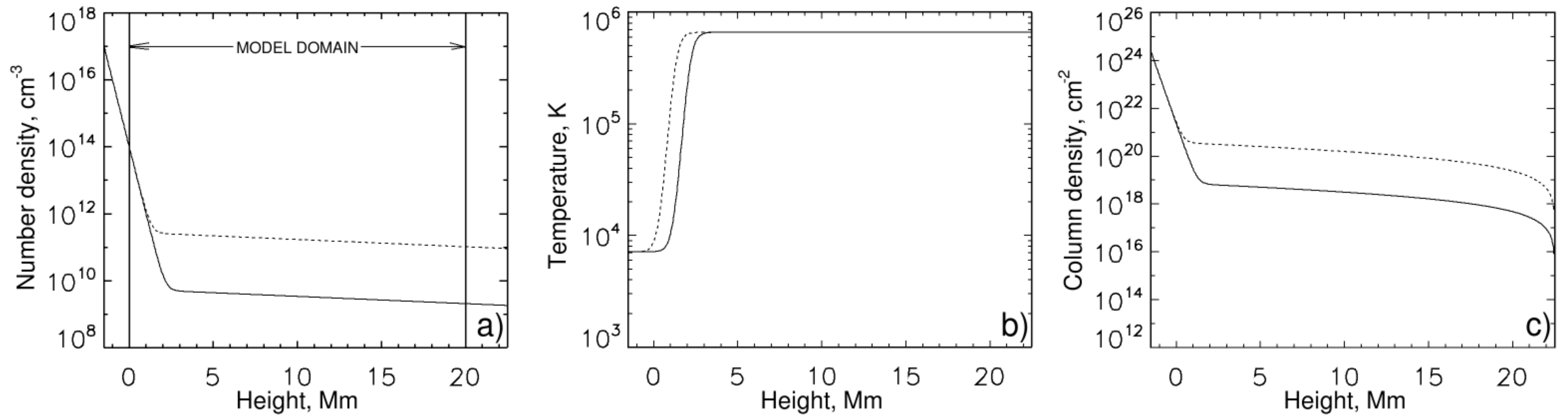}}
\caption{Initial density (panel a), temperature (panel b), and column density (panel c) distributions. Solid lines correspond to the low-density case (Model A); dotted lines correspond to the high-density case (Model B).}
\label{f-atmo}
\end{figure*} 

\citet{gobr11,gobr12} show that the energy release during reconnection in a twisted loop could provide a more convenient model of 
particle acceleration when compared to the standard model. The main benefit 
is that particle energization is distributed within the loop, 
rather than localized in the upper corona and hence, the energy losses due to various transport effects (collisions, return current) 
are reduced. Indeed, it is easier to accelerate a larger number of particles in a large-scale distributed region rather than in a compact volume, and hence, this scenario may have important consequences for the particle number problem 
\citep[see e.g.][]{brow76}. 
\citet{gore12} also demonstrated that reconnection in twisted loops with magnetic field convergence near footpoints 
could provide a configuration for particle re-acceleration in the lower corona and chromosphere \citep{broe09}. 
They also show that the
developed model can explain some features of the variation in hard X-ray sources recently observed with RHESSI \citep{kone11}. 

\citet{gobr12} used the model developed in \citet{hooe09} and \citet{gobr11} to study particle acceleration in twisted 
loops using the test-particle approach. This study recently has been extended by \citet{gore12} to incorporate collisions and 
magnetic field convergence. The obtained time-dependent particle distributions are used to predict 
the structure of hard X-ray sources.

In the present paper we develop a model of energy release in an unstable twisted coronal loop, focusing both on thermal and 
non-thermal components. The model combines 3D MHD simulations and test-particle calculations of electron and proton trajectories in presence of Coulomb collisions. 
The key new feature of this model are the large-scale curvature of the twisted magnetic fluxtube, and stratification of 
the ambient atmosphere. The latter affects the MHD part of the simulations through density-dependent anomalous resistivity, 
as well as particle acceleration due to Coulomb scattering of high-energy electrons and, to a lesser extent, protons. In the present 
paper we also calculate hard X-ray bremsstrahlung emission, enabling direct comparison of our model with observations.

%%%%%%%%%%%%%%%%%%%%%%%%%%%%%%%%%%%%%%%%%%%%%%%%%%%%%%%%%%%%%%%%%%%%%%%%%%%%%%%%%%%%%%%%%%%%%%
\section{Evolution of kinking twisted coronal loops}

In this section, we consider the evolution of a curved twisted coronal loop using a MHD approach. The method and model set-up
are described in Sect.~\ref{mhd-meth}. The derivation of the configuration with twisted fluxtube using ideal MHD is described
in Sect.~\ref{mhd-rota}, and the evolution of twisted loop after kink instability is described in the Sect.~\ref{mhd-kink}.

\subsection{The model and main equations}\label{mhd-meth}

Consider a three-dimensional domain with the dimensions $x=[-x_{\max};+x_{\max}]$, $y=[-y_{\max};+y_{\max}]$ and 
$z=[0;z_{\max}]$. The evolution of magnetic field $\vec{B}$, electric field $\vec{E}$, current density $\vec{j}$, plasma 
velocity $\vec{v}$, density $\rho$ and specific thermal energy $w$ with time $t$ can be described by the standard set of 
resistive MHD equations:

\begin{eqnarray}
\frac{\partial \rho}{\partial t} &=& - \vec{\nabla} \cdot (\rho \vec{v}) \label{mhd-rho}\\
\frac{\partial \vec{v}}{\partial t} &=& - (\vec{v} \cdot \vec{\nabla}) \vec{v} - \frac 1{\rho} \vec{j} \times \vec{B} - \frac 1{\rho}\vec{\nabla}p + \rho \vec{g} \label{mhd-velo}\\
\frac{\partial w}{\partial t} &=& - (\vec{v} \cdot \vec{\nabla})w - (\Gamma-1)w\; \vec{\nabla} \cdot \vec{v} + \frac{\eta}{\rho}j^2 +S_{other} \label{mhd-nrg}\\
\frac{\partial \vec{B}}{\partial t} &=& - \vec{\nabla} \times \vec E \label{mhd-magn}\\
\vec E &=& - \vec{v} \times \vec{B} + \eta \vec{j} \label{mhd-ele}\\
\vec{j} &=& \frac 1{\mu_0} [\vec{\nabla} \times \vec{B}]\label{mhd-curr} \\
p &=& (\Gamma -1 ) \rho w. \label{mhd-gas}
\end{eqnarray}
Here, $\mu_0$ is the magnetic permeability of the vacuum and $\eta$ is resistivity. The source term $S_{other}$ accounts for all other sources of heating, mostly due to numerical dissipation and shock viscosity used in the MHD simulations \citep[see][]{arbe01}.
The specific heat ratio is $\Gamma=5/3$. The temperature can be derived from the specific energy as 
$T=\mathcal{M}(\Gamma-1)/\mathcal{R}\;w$. Here, $\mathcal{R}$ 
is the universal gas constant, and $\mathcal{M}$ is the average molar mass equal to $\mathcal{M}=0.5 m_p N_A$, 
corresponding to fully ionised hydrogen. The gravitational acceleration, $\vec{g}=- g \vec{z}$, 
is assumed to be constant in the
whole domain. The last term in Eq.~\ref{mhd-nrg} accounts for all other possible sources of heating. In the present simulations, 
it is an artificial viscous dissipation used to stabilise the solution.

The set of Eqns \ref{mhd-rho}-\ref{mhd-gas} is solved numerically using Lagrangian remap MHD code LARE3D by \citet{arbe01}. 
The numerical grid of 256$\times$256$\times$512 elements (along the X-, Y-, and Z-axis, respectively) is uniform 
in each direction. (The 
resolution is higher in the vertical direction to reduce the error due to density stratification.)

The lower boundary of the domain is used as a driver to create a configuration with a twisted rope, as discussed in the 
Sect.~\ref{mhd-rota}. Conditions for the four side boundaries are based upon the assumption that the plasma is 
stationary ($\vec{v}=0$), and all the remaining parameters are constant across a boundary ($\frac{\partial \vec{B}}{\partial n} =0$, 
$\frac{\partial w}{\partial n} =0$, $\frac{\partial \rho}{\partial n} =0$, where $\frac{\partial}{\partial n}$ denotes a 
gradient across a boundary). 
All the parameters ($\vec{v}$ and $\vec{B}$, $w$ and $\rho$) are constant through the upper boundary. In addition, gravitational 
acceleration is set to zero at the upper boundary to maintain its stability \citep[see e.g.][]{arce10}.

Initially, there is a plain-parallel gravitationally stratified atmosphere. The initial density distribution is constructed so that 
it is similar to that in empirical models \citep[e.g.,][]{vere81}. At the photospheric level, the density is expected 
to be around $1.67\times 10^{-4}$ kg~m$^{-3}$, and the temperature is $\sim 6000$ K. In the model chromosphere, the density decreases exponentially 
by a factor of $100$ for every $1$ Mm. At $z \approx 2-3$~Mm, the temperature drastically increases to $\sim 1$ MK. In the model corona, 
the scale height becomes substantially larger, so that the density decreases by a small factor, while the temperature is nearly constant. 
The initial density is defined using the following formula:

\begin{equation}\label{eq-initatm}
\rho(z) = \rho_1 \exp \left(-\frac{z-z_{ch}}{z_1}\right)+\rho_2 \exp \left(-\frac{z-z_{ch}}{z_2}\right),
\end{equation}
and the corresponding initial pressure can be calculated from
\begin{equation}
\frac{dp}{dz} = \rho g.
\end{equation}
The temperature is then derived from the gas law.
Parameters $\rho_1$, $z_1$, $\rho_2$, and $z_2$ are chosen so that the initial density and temperature profiles satisfy the criteria described 
above, while $z_{ch}$ is used to shift the profile in a vertical direction. In the present simulations, we use two initial atmospheres: 
low-density with the coronal density corresponding to quiet corona ($\sim 10^9$~cm$^{-3}$) (Model A thereafter), and 
high-density with the coronal density 
corresponding to cases of extremely high density above active regions ($\sim 10^{10}$~cm$^{-3}$) (Model B thereafter)
\citep[see e.g.][]{trie08}. The parameters 
for Eq.~\ref{eq-initatm} are given in Table~\ref{t-atmo}, and the density and temperature for Models A and B are shown in 
Fig.~\ref{f-atmo}. In addition, Fig.~\ref{f-atmo} shows the column density defined as
\begin{equation}
\xi(z) = \int \limits_{z}^0 \frac {\rho(z)}{m_p} dz,
\end{equation}
where $m_p$ is the proton mass.

\begin{table*}
\label{t-atmo}
\caption{Parameters for the initial density distribution defined by Eq.~\ref{eq-initatm}}.
\centering
\begin{tabular}{l l l l l l}
\hline 
 & $\rho_1$, kg m$^{-3}$ & $z_1$, m & $\rho_2$, kg m$^{-3}$ & $z_2$, m & $z_{ch}$, m \\
\hline
\begin{tabular}{l} 
Model A \\
Model B
\end{tabular}%
& $1.7\times 10^{-4}$ & $2.2\times 10^5$ & %
\begin{tabular}{l} 

$1.0\times 10^{-11}$\\ 
$5.0\times 10^{-10}$
\end{tabular} %
& $2.0\times 10^6$ & $1.5\times 10^5$\\
\hline
\end{tabular}
\end{table*}

By introducing the scale length $L_0$, characteristic magnetic field $B_0$ and characteristic plasma density $\rho_0$ one can derive 
the dimensionless variables for all used physical parameters. Thus, $\vec{B}^*=\vec{B}\; B_0^{-1}$, $\vec{j}^*=\vec{j}\; \mu_0 L_0 B_0^{-1}$,
$\vec{E}^*=\vec{E}\; \mu_0^{1/2} \rho_0^{1/2} B_0^{-2}$, $\rho^*=\rho\; \rho_0$, $\vec{v}^*=\vec{v}\; \mu_0^{1/2} \rho_0^{1/2} B_0^{-1}$,
$t^*=t \; B_0 \mu_0^{-1/2} \rho_0^{-1/2} L_0^{-1}$, $w^*= w \; \mu_0 \rho B_0^{-2}$, and $p^*= p \mu_0 B_0^{-2}$, and
$g^* = g \mu_0 \rho_0 L_0 B_0^2$, where starred variables denote dimensionless physical parameters. The position vector and all 
the length variables are scaled as $s^*= s L_0^{-1}$. The dimensionless resistivity can be derived as 
$\eta^* = \eta \; \mu_0^{-1/2} \rho_0^{1/2} L_0^{-1} B_0^{-1}$. Later in the paper we refer to the dimensionless 
parameters (dropping the star symbol) everywhere, except for the cases where units are given explicitly. In the present 
simulations, we use the following scaling parameters: $L_0 = 10^6$ m, $B_0 = 4\times 10^{-3}$ T, and
$\rho_0 = 3.3\times 10^{-12}$~kg m$^{-3}$, yielding the characteristic velocity $v_0=1.95\times 10^6$ m s$^{-1}$ and 
time $t_0=0.53$ s.

\subsection{Derivation of configuration with twisted fluxtube}\label{mhd-rota}

A magnetic field configuration with a twisted, bended fluxtube is constructed by adding a twist to an initially potential 
magnetic field. 

The initial configuration is the potential field of two magnetic monopoles located outside the numerical domain:

\begin{equation}\label{eq-inifield}
\vec{B}(t=0)=B_1 L_0^2 \left(\frac{\vec{r}-\vec{m}_1}{|\vec{r}-\vec{m}_1|^3} - \frac{\vec{r}-\vec{m}_2}{|\vec{r}-\vec{m}_2|^3}\right),
\end{equation}
where $\vec{r}$ is the position vector, and $\vec{m}_1$ and $\vec{m}_2$ are the positions of the monopoles: 
$\vec{m}_1=[0;a;-h]$ and $\vec{m}_2=[0;-a;-h]$. The depth 
of the ``monopoles'', $h$, determines the convergence of the magnetic field towards footpoints. In our simulations, $a=6.4 L_0$ and $h=0.2 L_0$, providing the convergence ratio (the magnitude of magnetic field at the footpoint to that at the loop-top) of 
$\sim 10$.

\begin{figure}
\centerline{\includegraphics[width=0.4\textwidth]{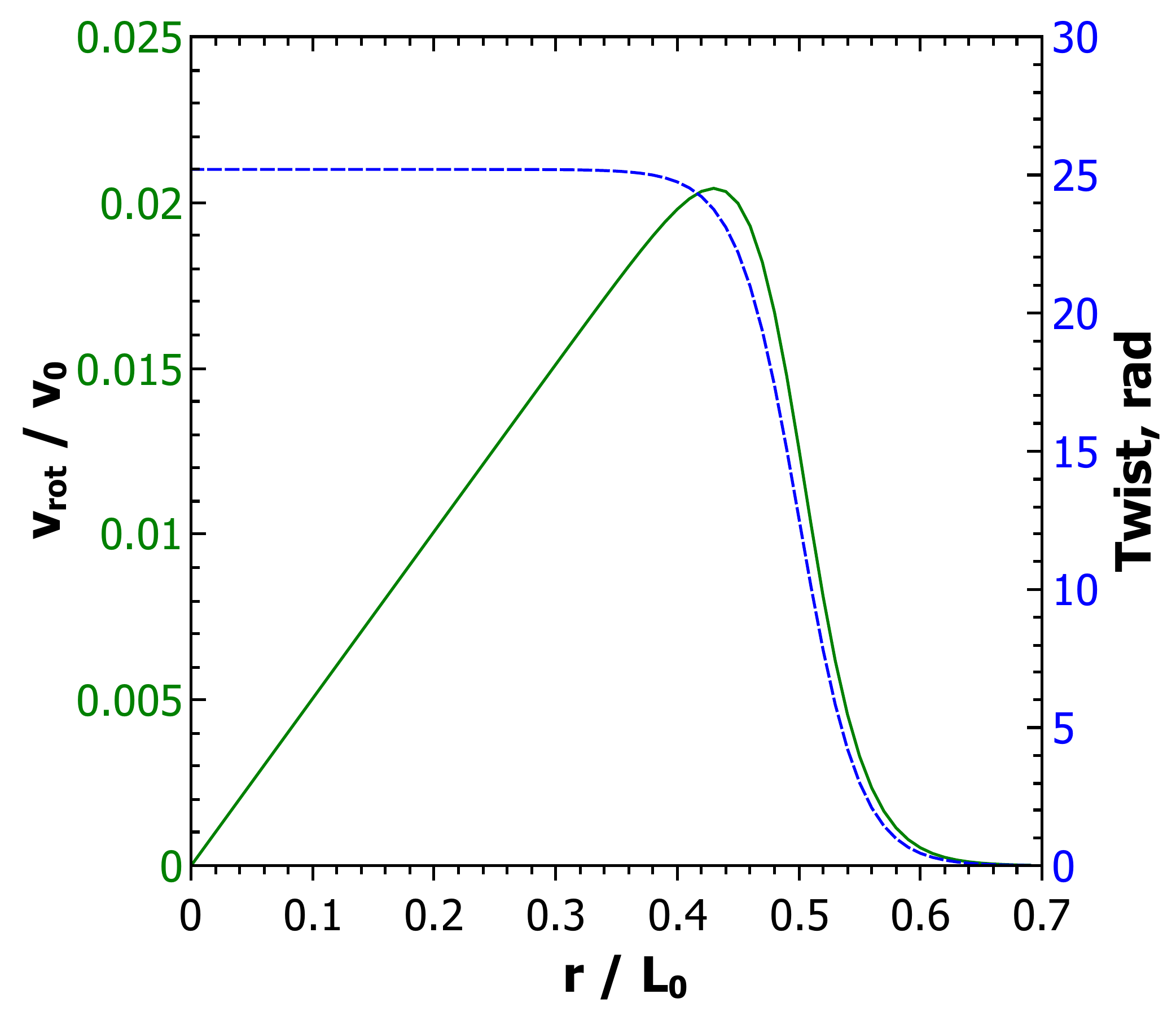}}
\caption{Maximum rotational velocity (solid line) and final twist angle (dashed line) as functions of the distance from the centre 
of rotation.}
\label{f-twist}
\end{figure}

The helicity is injected into the system by applying rotation at the footpoints. 
Then we follow the evolution of the magnetic field
using ideal MHD. The centres of rotation are the points with the maximum magnetic field at the lower boundary:
$x=0$ and $y\approx \pm 6.4 L_0$. Rotational velocity corresponds to the solid rotation case (see Fig.~\ref{f-twist}), where 
$v_{rot} \sim r$ inside the footpoint radius $R=0.5 L_0$: 
\begin{eqnarray}
v_{rot}(r)&=& \omega_{twist}r \times \nonumber \\
&& \left[1-\tanh \left(\frac{r-R}{0.05 L_0}\right)\right]\frac t{t_{twist}}\exp \left(-\frac t{t_{twist}}\right),
\end{eqnarray}
where $\omega_{twist}$ and $t_{twist}$ are the characteristic angular velocity and time, respectively.

The rotation is applied through $v_x$ and $v_y$ components at the lower boundary, while the vertical 
velocity is zero. The magnetic field components are kept constant across the lower boundary ($\partial \vec{B}/\partial z =0$), 
along with the specific internal energy ($\partial w/\partial z =0$). The density at the lower boundary is kept constant,  $\rho(z=0,t)=\rho(z=0,t=0)$, and similar to
the upper boundary, the gravitational acceleration is set to zero here to maintain boundary stability. 
In all the experiments, we have used $\omega_{twist}=0.1$ rad/s and $t_{twist}=250 t_0$. Hence, the rotational velocity is 
low, $v_{twist} \sim 10^{-2} v_0$, and the field structure, effectively, goes through a sequence of quasi-equilibrium states. 
After $t_{twist}\sim 10^3 t_0$, the total twist is about $8\pi$, and the magnetic configuration is nearly force-free 
but apparently kink-unstable. The obtained field geometry is similar to that
in \citet{gore12}, except we now have a more realistic bended loop and gravitationally stratified atmosphere.

\begin{figure}
\centerline{\includegraphics[width=0.4\textwidth]{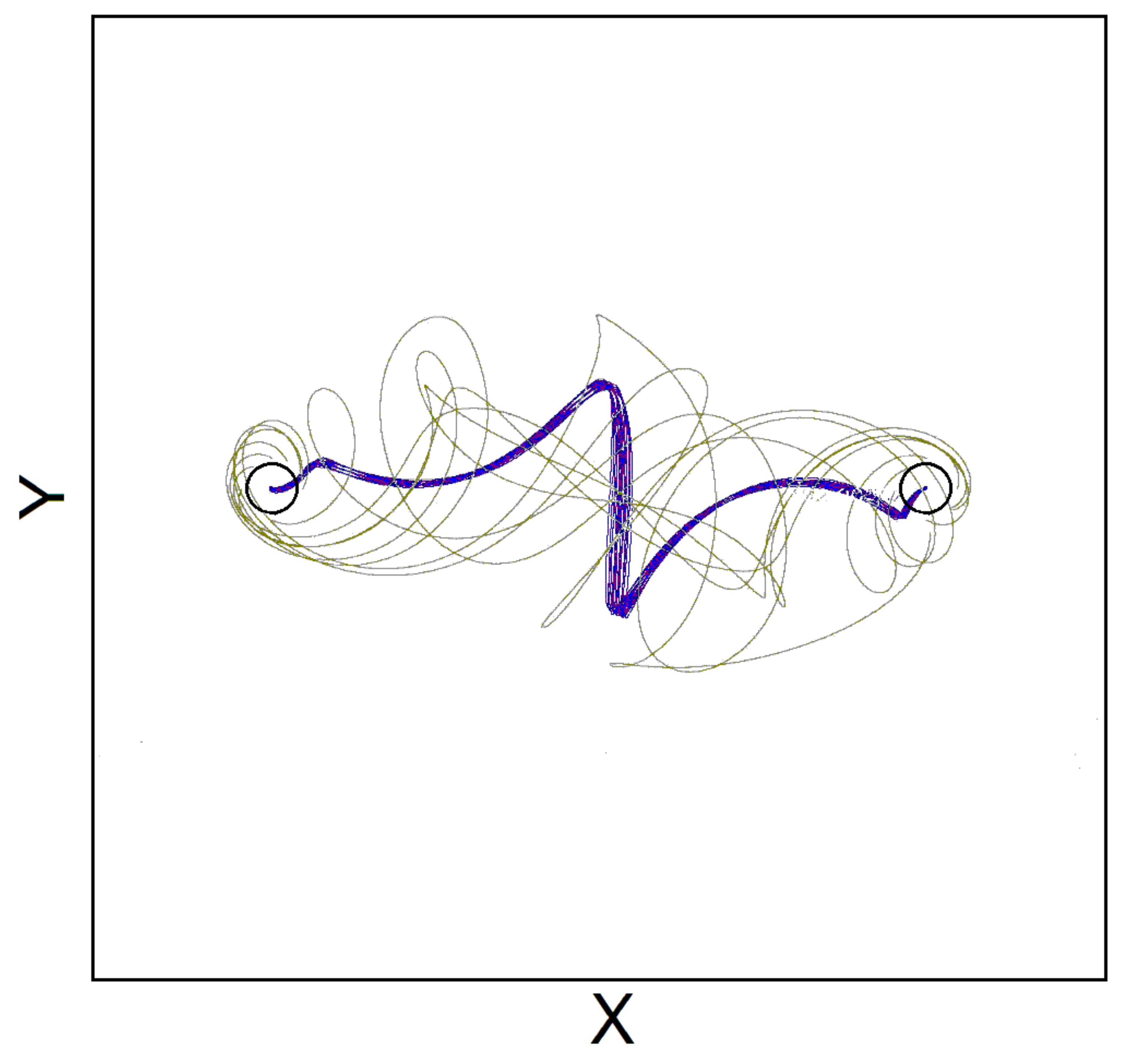}}
\caption{Magnetic field lines during kink instability development: view from the top at $t\approx 800 t_0$. The blue colour denotes the field lines connecting the footpoint centres; green colour denotes an ambient field.}
\label{f-sigma}
\end{figure}

The evolution of the models during and after the onset of kink instability is considered using resistive MHD, as described 
in the Sect.2.3.

\subsection{Magnetic reconnection in twisted coronal loops}\label{mhd-kink}

The kink instability occurs 
approximately at $t=920 t_0$ and $t=950 t_0$ in the Model A and Model B, respectively. In both cases, the $m=1$ kink mode dominates, resulting in the sigmoid structure (Fig.~\ref{f-sigma}). 

\begin{figure*}
\centerline{\includegraphics[width=0.75\textwidth]{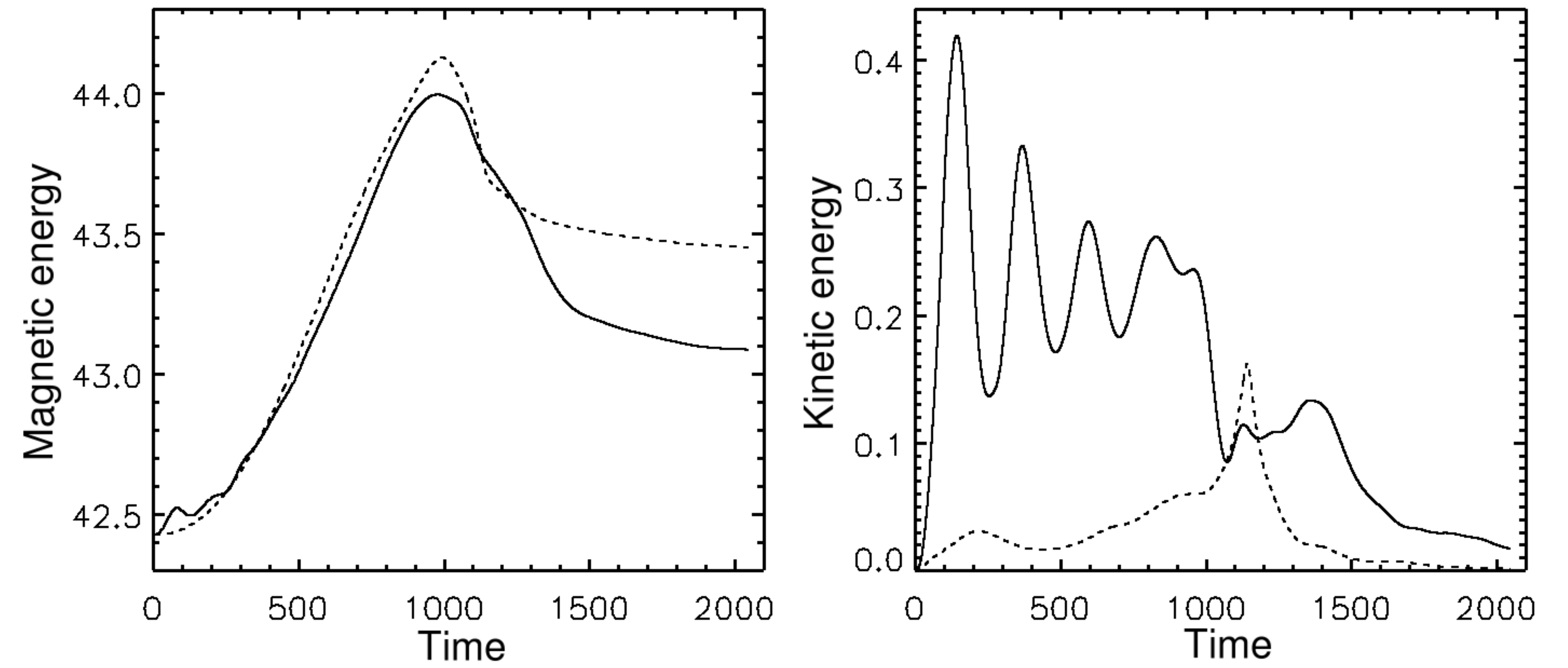}}
\caption{Total magnetic energy and total kinetic energies as functions of time for the low-density (solid lines) and the
high-density models (dashed lines). Time is measured in $t_0$ units, while the energies are measured in 
$\frac 12 \mu_0^{-1} B_0^2 L_0^3$.}
\label{f-mhdnrg}
\end{figure*}

\begin{figure*}
\centerline{\includegraphics[width=0.8\textwidth, height=0.9\textheight]{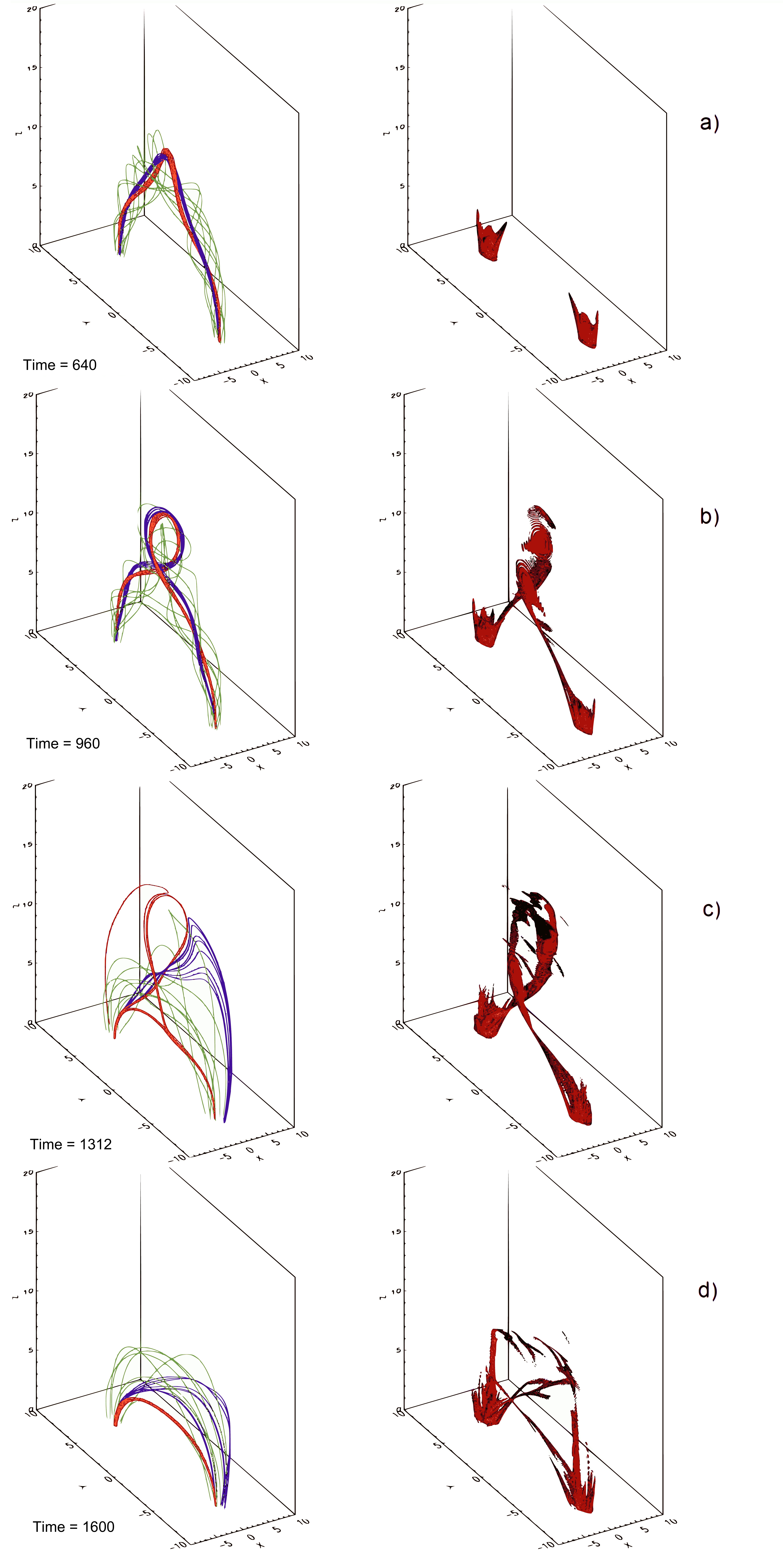}}
\caption{Selected magnetic field lines (left panels) and current density iso-surfaces ($j=0.5$) (right panels) during magnetic
reconnection in the Model A (low-density case). Different colours are used for magnetic field lines to demonstrate the change of connectivity. Blue lines originate almost at the centre of the footpoint  at $y\approx 6.4$; red lines originate 
almost at the centre of the footpoint ($y\approx -6.4$); other lines belonging to the twisted fluxtube are 
shown in green. The corresponding times are shown at the lower left corners.}
\label{f-recon}
\end{figure*}

The simulations are switched from an ideal to resistive MHD regime at $t=896 t_0$. The resistivity consists of two components: 
uniform and constant background resistivity $\eta_{back}$ and non-uniform current-driven anomalous resistivity $\eta_{cr}$. The 
need to specify 
some functional form for anomalous electric resistivity (or magnetic diffusion coefficient) is an intrinsic flaw in MHD 
simulations of magnetic reconnection; the resistivity in most numerical models is arbitrary to some extent. Many studies, however,
attempt to use some physically motivated form of anomalous resistivity, most often assuming that it is caused by the 
development of 
ion-acoustic turbulence \citep[see][]{ugai92,uzde03,bare11}. In the present simulations, we adopt similar approach and use  
current-driven anomalous resistivity. The total resistivity is defined as 
\begin{equation}\label{mhd-eta}
\eta=\eta_{back}+\eta_{cr} {\rm sign}(j-j_{cr}).
\end{equation}
The critical current $j_{cr}$ can be derived from the criteria $v_{drift} > v_{cr}$, which is normally used in such simulations 
\cite[e.g.][]{klie00}. Here, $v_{drift}=j/(en)$ is the electron drift velocity and the critical velocity equal 
to the sound speed $v_{cr}=\sqrt{\Gamma \frac p \rho}$. Hence, the expression for a critical current can be written as
\begin{equation}
j_{cr} = en\sqrt{\Gamma \frac p \rho}, \nonumber
\end{equation} 
and by using the scaling parameters introduced above, this formula can be reduced to a more convenient form
\begin{equation}
\frac{j_{cr}}{j_0} = \sqrt{\frac \gamma 2} \beta_g \frac {L_0}{R_{L,\;p}} \frac{\rho}{\rho_0}, \nonumber
\end{equation} 
where $\beta_g = \frac{2 \mu_0 nkT}{B_0^2}$ is the gas to magnetic pressure ratio and 
$R_{L\;p}= \frac{m_p}{e B_0} \sqrt{\Gamma \frac p \rho}$ is the Larmor radius of a thermal proton.
The problem, however, is that in global MHD models the current density is 
limited by the numerical grid resolution. In our simulations the grid resolution is 
$\delta L = L_0/256 \approx 4\times 10^{-3} L_0$ or $\sim 4$~km in real terms, while it is 
widely accepted that a realistic thickness of a RCS is
about the Larmor radius of the proton $R_{L,\;p}$, which is $\sim 1$~m in the corona. Hence, the grid resolution 
is by a factor $\sim 10^3-10^4$ larger than realistic RCS thickness, which means that the real current densities should be 
$\sim 10^3-10^4$ higher. To overcome this problem, we introduce
a correcting factor of $\delta L/R_{L,\;p}=4\times 10^3$ for $j_{cr}$ resulting in the following form:

\begin{equation}
\frac{j_{cr}}{j_0} = \sqrt{\frac \Gamma 2} \beta_g \frac {L_0}{\delta L} \frac{\rho}{\rho_0}, 
\end{equation} 
which brings the critical current just above the current densities before the kink instability.

The background resistivity is $\eta_{back}=10^{-6}$. This value is substantially higher than real classical resistivity in the corona 
(which would be about $10^{-9}$, if we consider our scaling parameters), but it is not expected to affect the model, as
the corresponding diffusion time is $t_{diff} \approx (\delta L)^2/\eta_{back} \approx  \sim 10^2 - 10^3 t_0$, 
where $\delta L$ is the grid step. As far as the value for the anomalous resistivity is concerned, it is more
difficult to determine a physically-justified $\eta_{cr}$. Therefore, the 
experiment with low-density atmosphere (Model A) has been repeated with several values of $\eta_1$: $10^{-4}$, $4\times 10^{-4}$,
$10^{-3}$ and $4\times 10^{-3}$. As in earlier MHD experiments with Lare code, we did not find any noticeable difference in the 
equilibrium formed after the reconnection \citep{gore11}. The main difference is the speed of relaxation: it is nearly 
proportional to 
$\eta_{cr}^{3/5}$. Additionally, it was found that dissipation due to numerical viscosity increases for higher $\eta_{cr}$ 
values. This 
can be explained by faster plasma flows in models with faster energy release. This effect was found to be especially significant 
in the experiment with $\eta_{cr} = 4 \times 10^{-3}$: the energy lost due to (artificial) viscous dissipation was comparable 
to Ohmic heating. Therefore, for particle trajectory calculations in Section 3 we use experiments with $\eta_{cr}=10^{-3}$.

Once the field goes unstable, the current
density drastically increases near loop-top and footpoints. The current density is high in these two regions 
due to different reasons: near footpoints strong currents appear due to magnetic field convergence, while
the current density near the loop-top becomes high because of the kink resulting in formation of magnetic rope. 
These regions, however, are not compact: they are extended along the loop legs and  
from time to time the footpoint regions get merged with the loop-top one. 
The increase in the current density 
above critical threshold switches on the anomalous resistivity, which, in turn, 
results in magnetic reconnection and dissipation of magnetic energy. This is qualitatively similar to the reconnection 
scenario in previous studies \citep{broe08,hooe09,gobr12}.

Just after the kink, the current is concentrated in a cylindrical structure following the loop legs. At the top of the loop, 
there is a current ring corresponding to the magnetic plasmoid. The current density is particularly high near the footpoints 
(due to magnetic field convergence) and in the ring at the top of the loop. As in previous experiments 
\citep[e.g.,][]{hooe09,gobr11,bare13}, the current gradually becomes filamentary and weakens. It almost disappears after 
$t \approx 1900 t_0$, apart from the footpoints, where it remains below $j_{cr}$ level. 

Total magnetic and kinetic energies for the MHD models are plotted versus time in Fig.~\ref{f-mhdnrg}. 
The magnetic energy variations are similar in both models. 
The difference in the maximum energy and the energy of the relaxed state is due to the
density-dependent anomalous resistivity current threshold. The latter is higher in the high-density case, and this is why 
the current density and magnetic energy can reach a higher value before the reconnection starts. Similarly, as the 
reconnection proceeds, the maximum current density in this case gets below the critical current threshold earlier, while the magnetic energy is still relatively high. As far as the kinetic energy is concerned, the low- and high-density cases are different
during the helicity injection phase. Both model loops demonstrate weak kink oscillations; however 
the frequency of oscillations and their amplitude are substantially smaller in the high-density 
case (Model B) than those in the low-density case (Model A). After the kink occurs 
(at approximately $t > 900\, t_0$), the kinetic energy variations are similar: they peak at $\sim 0.002$ of initial model 
energy and then gradually decay. The kinetic energy at this stage is associated largely with reconnection outflows. 

\begin{figure*}
\centerline{\includegraphics[width=1.0\textwidth]{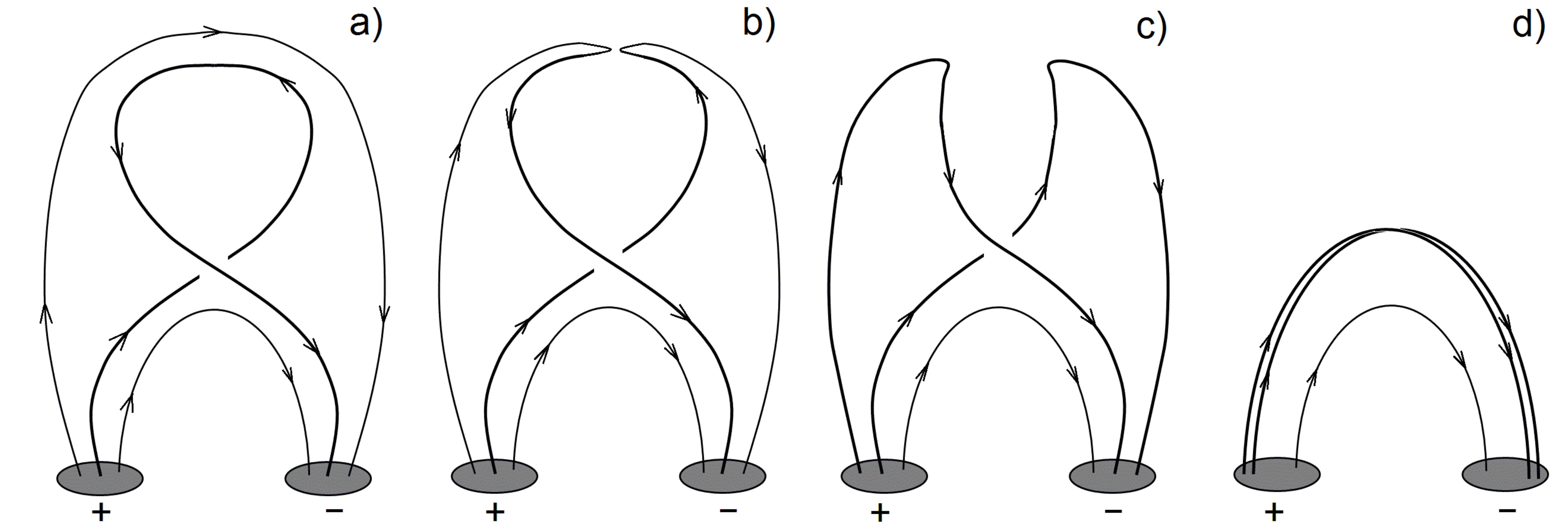}}
\caption{Schematic sketch showing change of connectivity during magnetic reconnection in the twisted loop.}
\label{f-reske}
\end{figure*}

In contrast with previous simulations of kink instability in coronal loops \citep{klie10}, we do not observe 
separation and emergence of the plasmoid.
The difference apparently appears because magnetic reconnection in our simulations occurs in a completely closed magnetic 
configuration, and the overlying ambient magnetic field lines are tied either to lower (photospheric) or side boundaries. 
The reconnection happens mostly at the very top of the kinking loop near the upper part of the plasmoid ring between 
the twisted field lines and the overlying ambient field lines. This results in the reduction of a twist and radial expansion 
of the loop, which is similar to the previous simulations of the initially cylindrical loop \citep{gobr11,gobr12}. The main 
two differences with these previous simulations are: (a) the process now, obviously, is not cylindrically symmetric and the current density varies strongly both along the loop and within its cross-section; (b) 
the absence of magnetic diffusion very close to the footpoints mean the two footpoints remain connected in the final 
equilibrium, and all the field lines remain connected to both footpoints (unlike in \citet{gobr11}). The latter is explained 
by the sketch in Fig.~\ref{f-reske}.

%%%%%%%%%%%%%%%%%%%%%%%%%%%%%%%%%%%%%%%%%%%%%%%%%%%%%%%%%%%%%%%%%%%%%%%%%%%%%%%%%%%%%%%%%%%%%%
\section{Electrons and protons in a twisted coronal loop}

In this section, we consider particle motion in reconnecting a twisted coronal loop based on the electric and magnetic fields, and 
plasma density obtained in the MHD simulations (see Sect.~\ref{mhd-kink}). The methodology is explained in Sect.~\ref{tp-meth}, 
which followed by a description of the trajectories (Sect.~\ref{tp-traj}), energy spectra (Sect.~\ref{tp-ensp}), and spatial distribution (Sect.~\ref{tp-spat}).

\subsection{Test-particle motion in presence of Coulomb collisions}\label{tp-meth}

Let us consider the motion of charged test-particles using the guiding-centre approximation with the particle trajectory described by the 
velocity parallel to magnetic field $v_{||}(t)$, drift velocity perpendicular to the magnetic field $\vec{u}(t)$, and 
Larmor gyration velocity $v_g(t)$. This approach is valid in this model, since the magnetic field is non-zero throughout the domain, 
and the proton and electron gyro-radii are much smaller than the characteristic scale $L_0$. 

The trajectories of changed particles in the presence of collisions can be described by the following set of equations:
\begin{eqnarray}
\frac {d \vec{r}}{dt} &=& {\vec{u}} + \frac{(\gamma v_{||})}\gamma \vec{b} \label{gca-pos}\\
\vec{u} &=& \vec{u}_E + %
\frac mq \frac {(\gamma v_{||})^2}{\gamma \kappa^2 B} [\vec{b} \times (\vec{b}\cdot \vec{\nabla})\vec{b}] + \nonumber \\%
&& \frac mq \frac \mu {\gamma \kappa^2 B} [\vec{b} \times \vec{\nabla} (\kappa B)]\label{gca-drift}\\
\frac{d (\gamma v_{||})}{dt} &=& \frac qm \vec{E}\cdot \vec{b} - \frac \mu \gamma (\vec{b} \cdot \vec{\nabla}(\kappa B))\nonumber \\%
&& \left[ v \frac{\delta \alpha}{\delta t}\right]_{coll} + \left[\alpha \frac {\delta v}{\delta t} \right]_{coll}\label{gca-vpara}\\
\frac{d\mu }{dt} &=& \left[\frac{v(1-\alpha^2)}B \frac{\delta v}{\delta t}\right]_{coll}%
-\left[\frac{\alpha v^2}B \frac{\delta \alpha}{\delta t}\right]_{coll}.\label{gca-mgm}
\end{eqnarray}
Here, $\vec{r}$ is particle position, $\mu$ is specific magnetic moment $\mu=\frac 12 v_g^2/B$, $\vec{u}_E$ is the 
E$\times$B-drift velocity $\vec{u}_E=\vec{E} \times \vec{B}/B^2$, and $\vec{b}$ is the magnetic field direction $\vec{b}=\vec{B}/B$.
Apart from the last two terms in Eq.~\ref{gca-vpara} and Eq.~\ref{gca-mgm}, this is the standard set of relativistic
guding-centre equations \citep{nort63}. The relativistic factors are defined as
\[
\gamma = \frac c {\sqrt{c^2-v_{||}^2-u^2-2\mu B}}
\]
and
\[
\kappa = \sqrt{1-u^2_E/c^2}.
\]

The collisional terms in the equations above are defined below by Eqs. (\ref{gca-collv}) and (\ref{gca-colla}). Let us 
discuss them and their numerical implementation in more details. In collisional terms, we ignore the relativistic factor $\gamma$,
since particles with a value of $\gamma$ noticeably greater than 1 are not significantly affected by Coulomb collisions.
The average collisional energy loss of a particle with 
mass $m$ moving with speed $v \gg v_{th} = \sqrt{\frac{k_B T}m}$ in fully ionised thermal plasma with temperature $T$ can be 
expressed as \citep[see e.g.][]{emsl78}:

\begin{equation}\label{gca-emslie}
\frac {d \mathcal{E}}{dl} = - 2 \pi e^4 \Lambda \frac m{m_e} \frac {n}{\mathcal{E}},
\end{equation}
where $\Lambda$ is the Coulomb logarythm. For non-relativistic particles, the kinetic energy
$\mathcal{E} = mv^2/2$ and Eq.~\ref{gca-emslie} can be rewritten as 
\begin{equation}\label{gca-collv}
\frac {d v}{dt} = -\mathcal{K} \frac {m_e}m \frac {n}{v^2}
\end{equation}
where $\mathcal{K}=4\pi e^4 \Lambda /m_e^2$ and $\Lambda \approx 20$.
The Eq.~\ref{gca-collv} yields variations in the full velocity $\delta v / \delta t$ for Eqs.~\ref{gca-vpara} and 
\ref{gca-mgm}. The full velocity in the right-hand side of  Eq.~\ref{gca-collv}
is calculated as $v=\sqrt{v_{||}^2 + 2 \mu B}$. Here the drift velocity is disregarded, since {\it (a)} it is normally 
comparable to or lower than the thermal velocity $v_{th}$, and {\it (b)} the main component of the drift, $\vec{u}_E$, is caused 
predominantly by bulk plasma motion. 

\begin{figure}
\centerline{\includegraphics[width=0.5\textwidth]{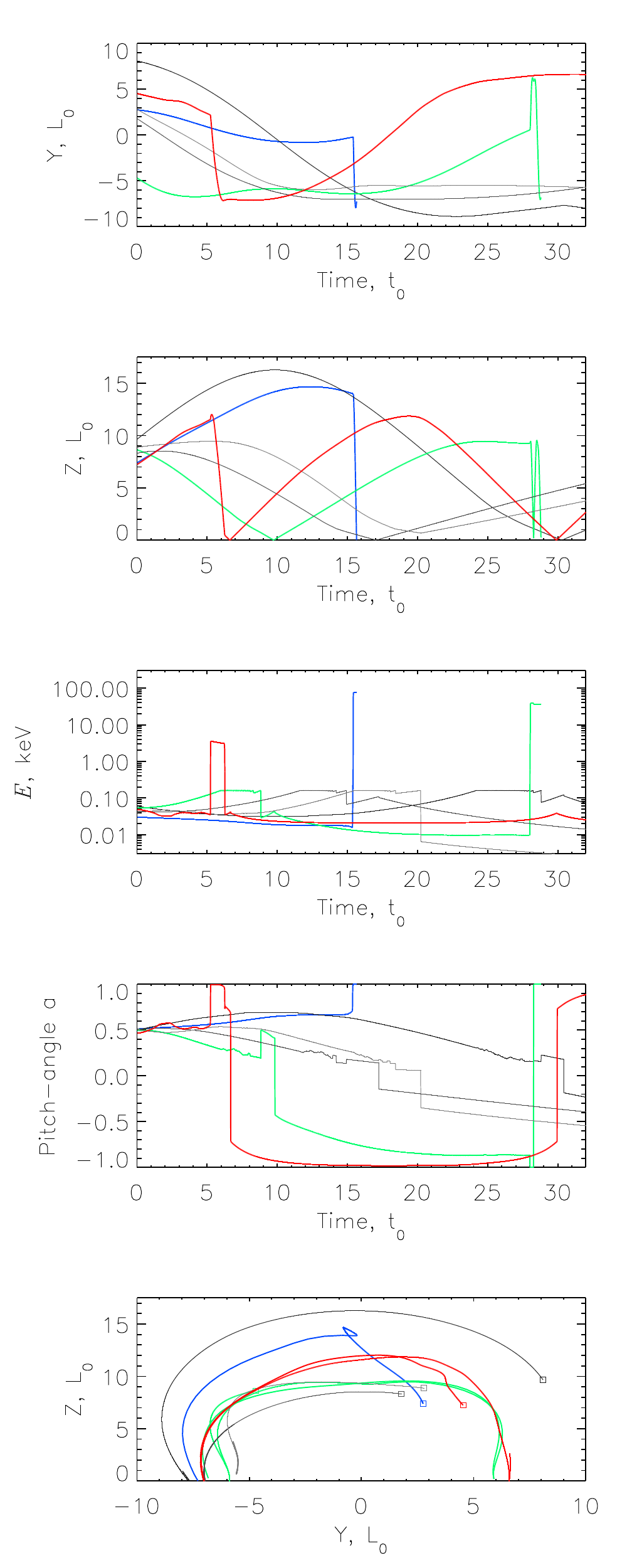}}
\caption{Typical electron trajectories, their energies and pitch-angles. Particles are injected with thermal energies at the locations denoted by squares on the 
lower panel. Blue, green, and red lines denote particles, which are non-thermal at some point, while grey lines denote particles,
which remain thermal throughout simulations. See text for more details.}
\label{f-etraj}
\end{figure}

\begin{figure}
\centerline{\includegraphics[width=0.5\textwidth]{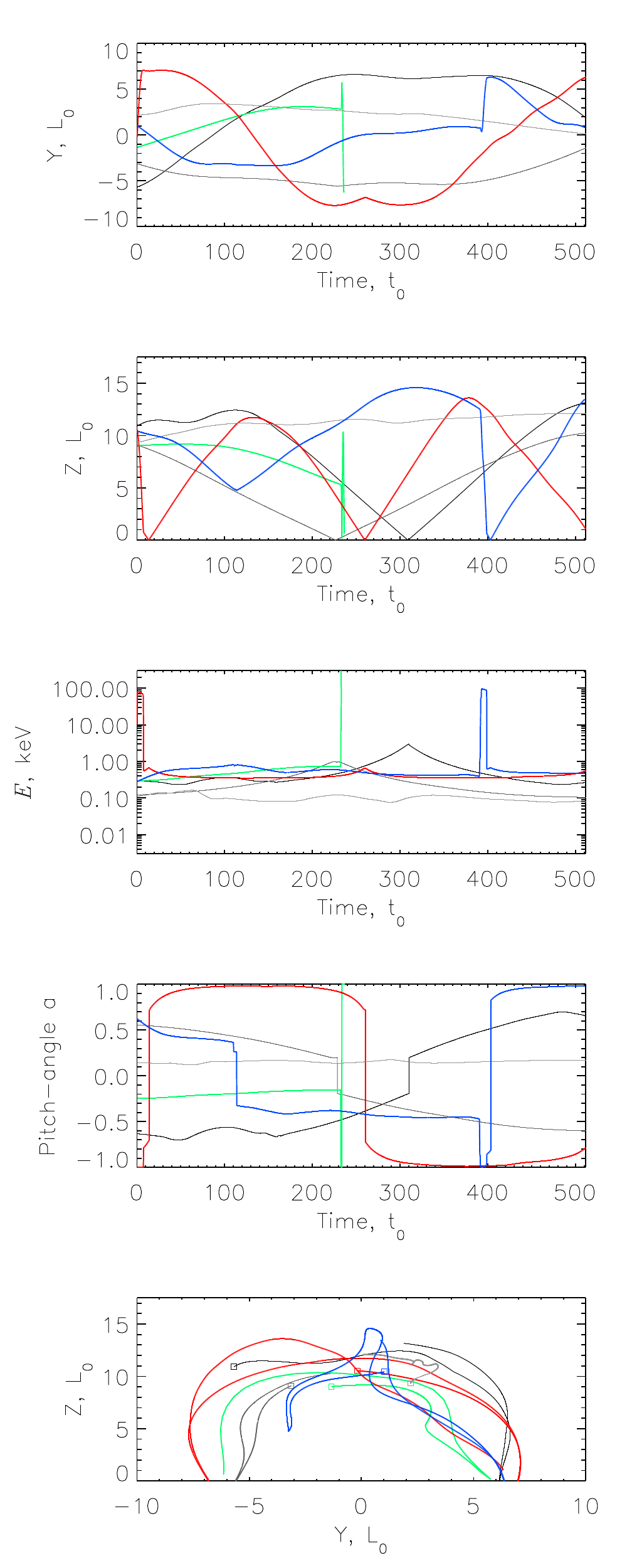}}
\caption{Typical proton trajectories, their energies and pitch-angles. Particles are injected with thermal energies at the locations denoted by squares on the 
lower panel. Blue, green, and red lines denote particles, which are non-thermal at some point, while grey lines denote particles,
which remain thermal throughout simulations. See text for more details.}
\label{f-ptraj}
\end{figure}

The pitch-angle distribution of high-energy particles in thermal plasma changes with time due to Coulomb collisions as
\begin{equation}\label{gca-collpa}
\frac {\partial f}{\partial t} = \mathcal{K} \frac m{m_e} \frac {n}{v^3} \frac {\partial}{\partial \alpha} 
\left((1-\alpha^2)\frac {\partial f}{\partial \alpha} \right ),
\end{equation} 
where the pitch-angle is defined as $\alpha=v_{||}/v$. In terms of individual particles, the pitch-angle diffusion 
represents random deflections with respect to a chosen direction. Let us evaluate the probability of a test-particle to be 
deflected. Equation \ref{gca-collpa} means that the fraction of particles changing its pitch-angle from $\alpha$ 
to $\alpha+\delta\alpha$ within the time interval $\delta t$ is 

\begin{equation}\label{gca-probab}
\Pi(v,\alpha, \Delta \alpha) = \delta t \mathcal{K} \frac m{m_e}\frac {n}{v^3} \frac{1-(\alpha+\Delta \alpha/2)^2}%
{\Delta \alpha^2}. \nonumber
\end{equation}
This, in turn, means the following: if individual particles are deflected by $\Delta \alpha$ with probability 
$\Pi(v,\alpha,\Delta \alpha)$ as per Eq.\ref{gca-probab}, then the pitch-angle distribution of the whole 
particle population should satisfy Eq.\ref{gca-collpa}.

In the numerical scheme used here, the pitch-angle diffusion is implemented through stochastic jumps of a particle pitch-angle by 
a fixed value of either $\Delta \alpha= 0.05$ or $\Delta \alpha= -0.05$ after every timestep $\delta t$ with the probability
of $\Pi(v,\alpha, \Delta \alpha)$ or $\Pi(v,\alpha, -\Delta \alpha)$, respectively (which means the pitch-angle remains the same 
with the probability $1-\Pi(v,\alpha, \Delta \alpha)-\Pi(v,\alpha, -\Delta \alpha)$). Hence, the variation in pitch-angle with time can be written as:
\begin{equation} \label{gca-colla}
\frac{\delta \alpha}{\delta t} = \Delta \alpha \Pi(v,\alpha, \Delta \alpha) - \Delta \alpha \Pi(v,\alpha, - \Delta \alpha)
\end{equation}
with probabilities $\Pi$ as per Eq.~\ref{gca-probab}.

In each numerical experiment, we calculate trajectories for $\sim 10^6$ test electrons and protons. Initially, particles are 
uniformly distributed within the simulation domain, have Maxwellian velocity distributions corresponding to the temperature of
$1$ MK, and are uniformly distributed with respect of the pitch-angle $\alpha$ from $-1$ to $1$. 

The calculations have been performed using the GCA code based on the second order Runge-Kutta scheme \citep{gore10,gore11}. 
In addition to the 
usual limitations on the integration timestep ($\delta t \ll \delta r/v$ and $\delta t \ll v/a$, where $\delta r$ is the grid step 
and $a$ is the acceleration), the collisional terms add another requirement: 
$\delta t \ll \mathcal{K} \frac m{m_e}\frac {n}{v^3} \frac{1-(\alpha+\delta \alpha/2)^2}{\delta \alpha^2}$,
which is necessary for $\Pi(v,\alpha, \delta \alpha) \ll 1$. (When $\Pi(v,\alpha, \delta \alpha) \sim 1$, the scheme
used for pitch-angle scattering calculations becomes unstable, and, obviously, the probability cannot be 
$\Pi(v,\alpha, \delta \alpha) > 1$.)

The time-dependent electric and magnetic fields and their spatial derivatives for the right-hand side of 
Eqs.~\ref{gca-pos}-\ref{gca-mgm} are taken from the resistive MHD simulations described in Sect.~\ref{mhd-kink}. The data
for each particle position is derived by linear interpolation of the data within four-dimensional $(x,y,z,t)$ cells 
from the ajacent grid points \citep[see][]{gore11,gobr11,gore12}. 
The domain boundaries are closed for thermal test-particles with $\mathcal{E}<1$~keV and open for higher energy particles. Each 
particle is followed until the end of simulations at $t=1800 t_0$ or until it leaves the domain through one of the boundaries. 

\begin{figure*}
\centerline{\includegraphics[width=0.7\textwidth]{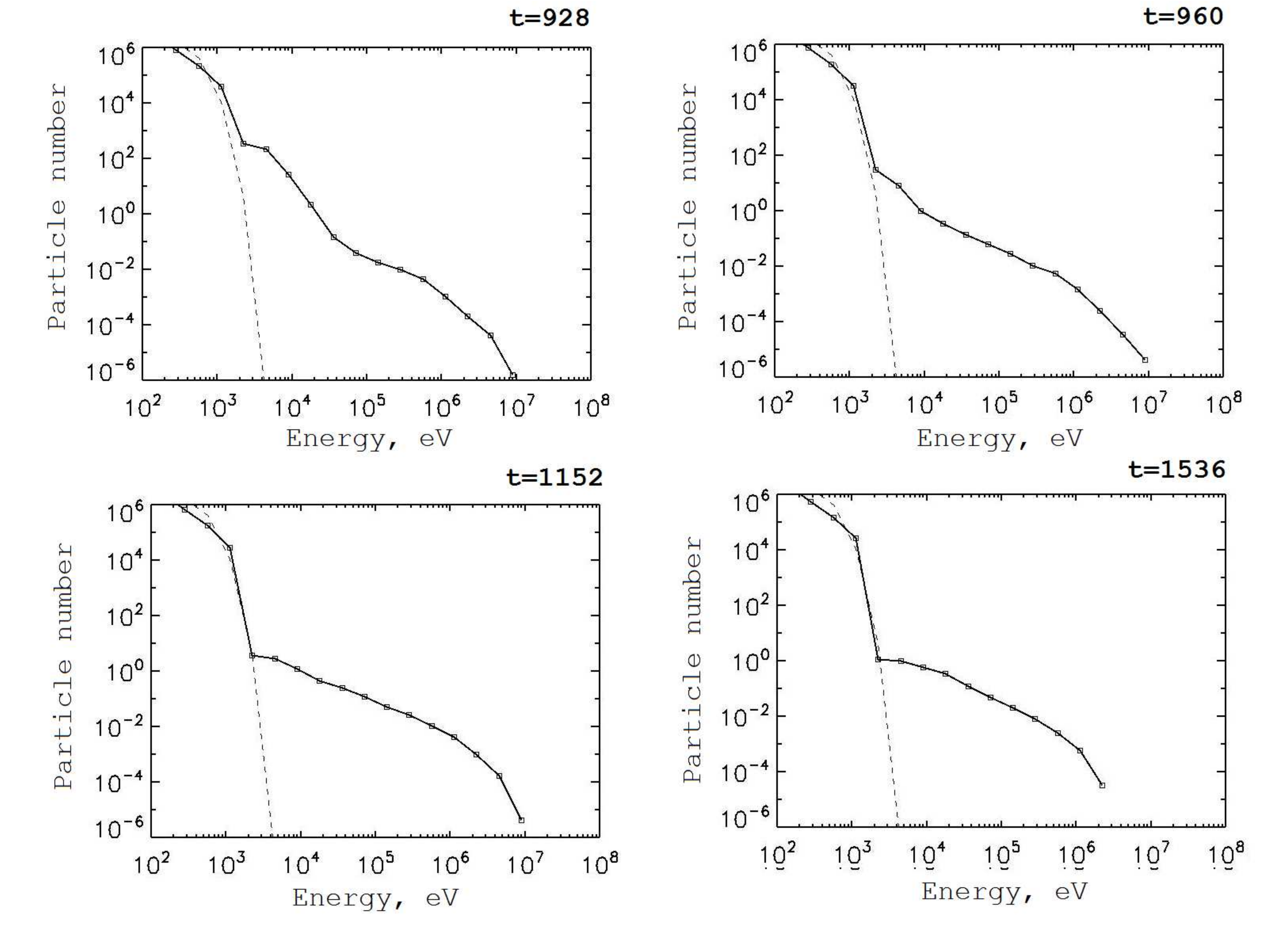}}
\caption{Electron energy spectra in the Model A. Times are shown above corresponding panels.}
\label{f-esplow}
\end{figure*}

\begin{figure*}
\centerline{\includegraphics[width=0.7\textwidth]{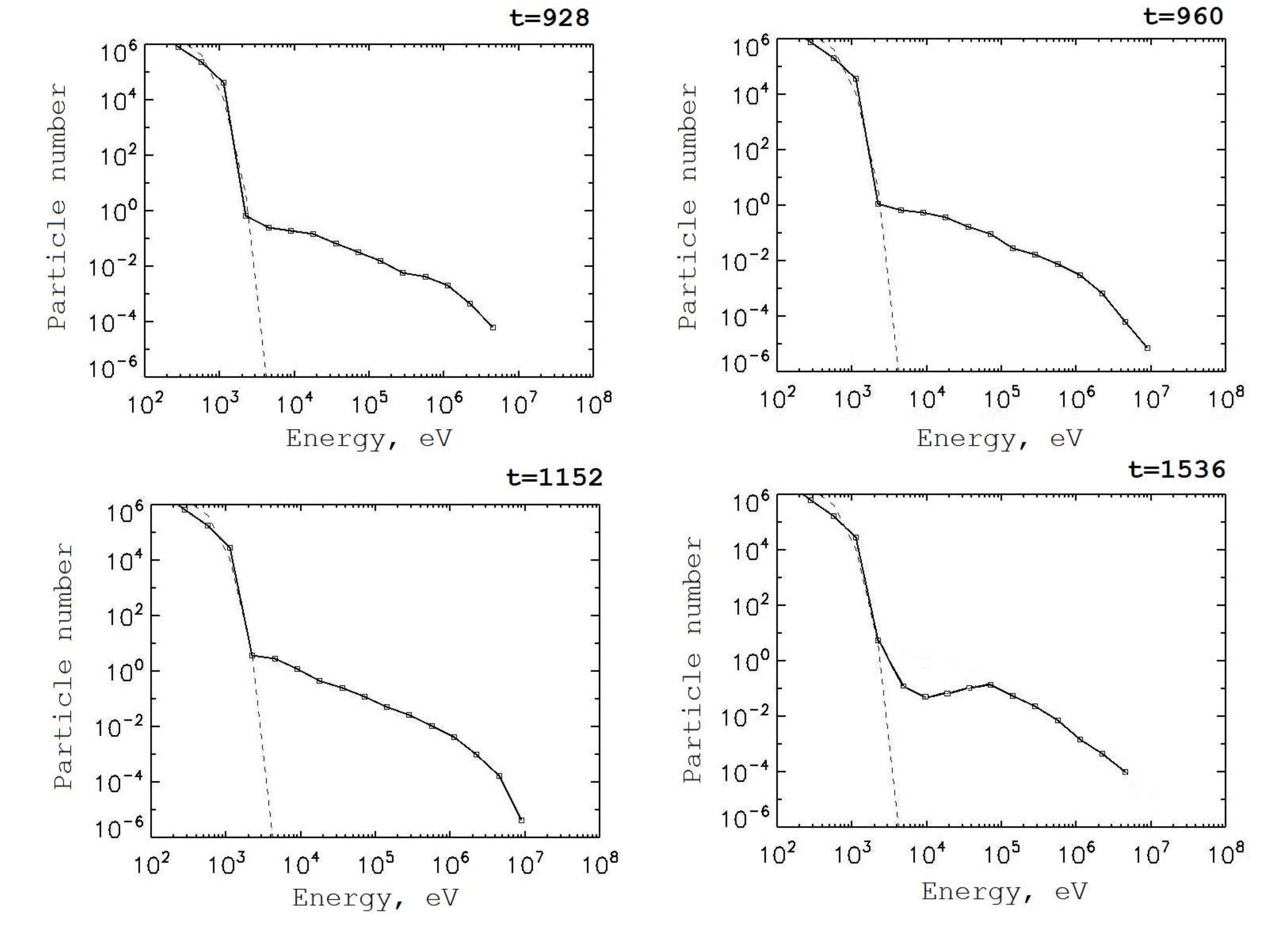}}
\caption{Electron energy spectra in the Model B. Times are shown above corresponding panels.}
\label{f-esphigh}
\end{figure*}

\begin{figure*}
\centerline{\includegraphics[width=0.7\textwidth]{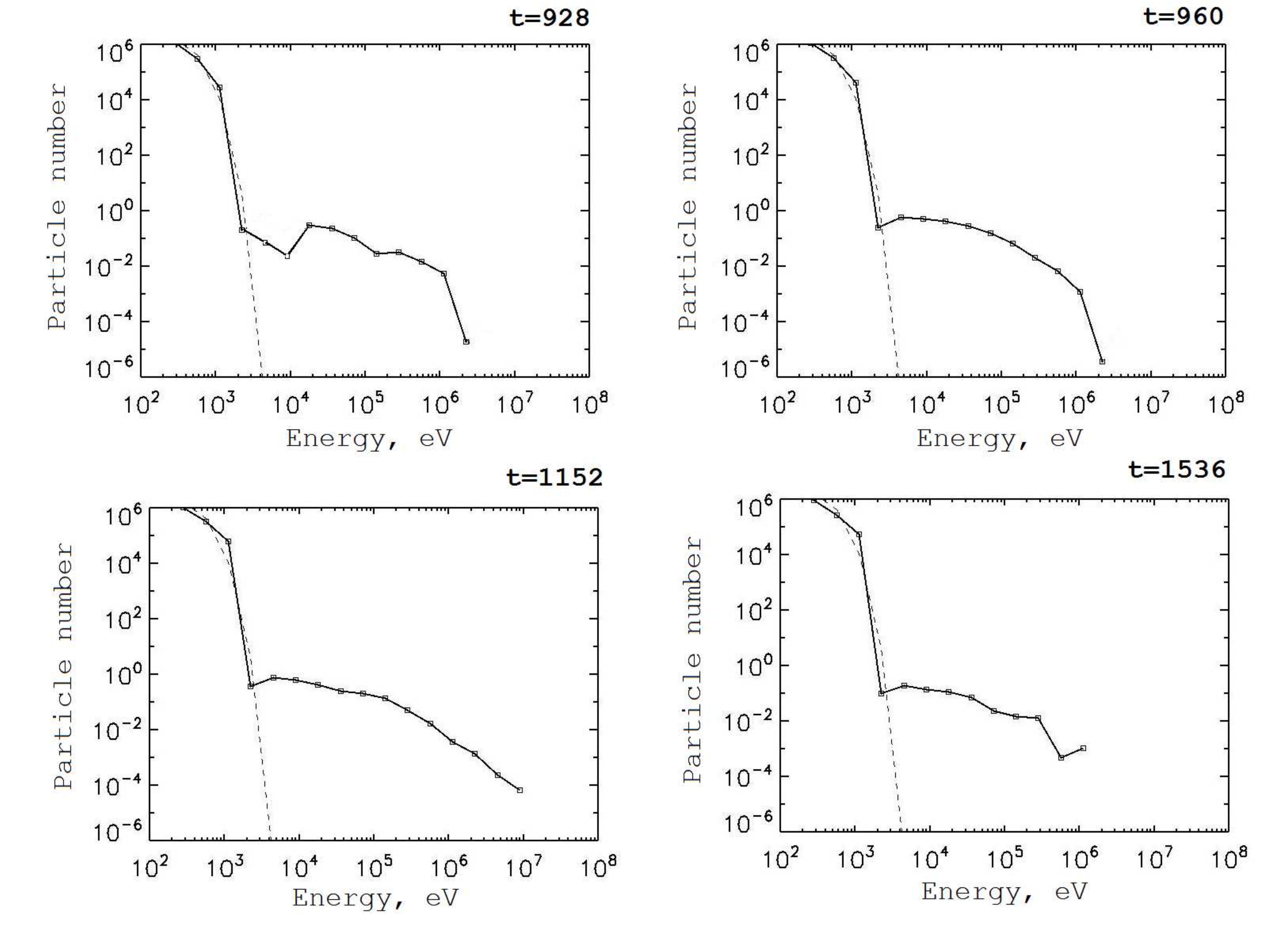}}
\caption{Proton energy spectra in the Model A. Times are shown above corresponding panels.}
\label{f-psplow}
\end{figure*}

\begin{figure*}
\centerline{\includegraphics[width=0.7\textwidth]{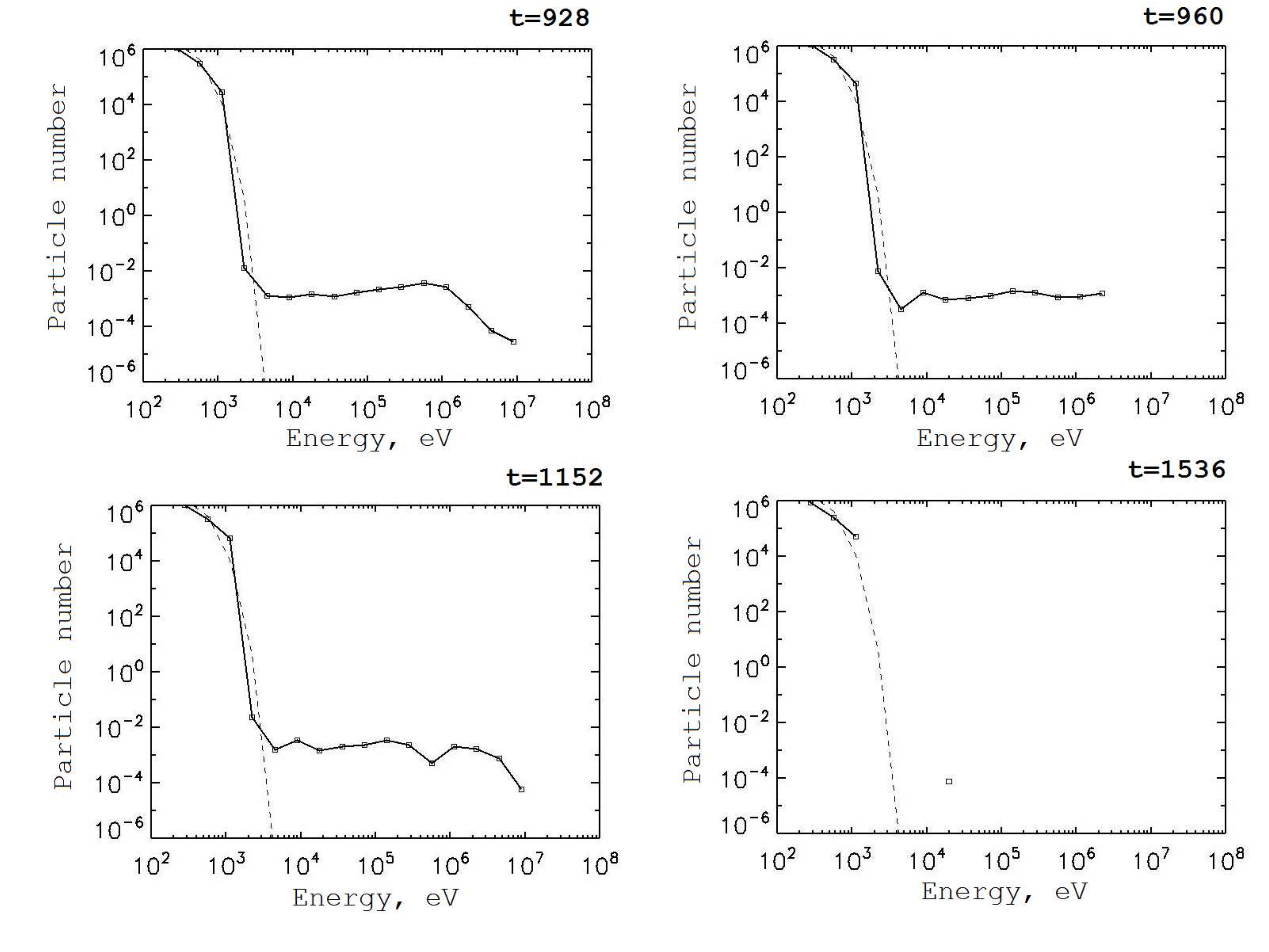}}
\caption{Proton energy spectra in the Model B. Times are shown above corresponding panels.}
\label{f-psphigh}
\end{figure*}

\subsection{Particle trajectories}\label{tp-traj}

Typical electron and proton trajectories are shown in the Figs. \ref{f-etraj} and \ref{f-ptraj}, respectively. In general, 
both species are very adiabatic and hence, behave  very similarly, The main difference, of course, is that protons are much 
(by factor of $\sim 43$) slower. As a result, their trajectories are smoother, since the effect of small-scale fluctuations (in 
the parallel electric field value, magnetic field curvature etc) is negligible due to time averaging.

The majority of particles remain in the thermal distribution; only a few particles ($\sim 5\%$) are accelerated beyond $1$~keV. This is comparable to typical acceleration efficiencies derived from observed hard X-ray emission.
This validates the use of a test-particle approach, as the effect of high-energy particles on the magnetic and electric field
should be considerably low.

Particles move predominantly along the magnetic field lines. Despite the connectivity 
changes, all the field lines of the twisted loop remain connected to both footpoints, and there is no open field.
As a result, during the reconnection most of high-energy particles remain in or around the twisted loop and precipitate 
towards one of the footpoints.

Particles accelerated by a parallel electric field in an almost collisionless corona have a very narrow pitch-angle distribution 
around $\alpha = \pm 1$, which means they are strongly collimated along the magnetic field \citep{gore11,gobr12}. However, they
get scattered due to collisions in the denser chromosphere, getting a wider pitch-angle distribution. As a result, there is  
a small but noticeable fraction of high-energy particles reflected by the converging field back to the corona.
Electrons with energies 
up to $\sim 100$~keV and nearly all the protons are thermalised before reaching the lower boundary. Hence, only a small fraction 
of energetic electrons (with $\mathcal{E} > 100$~keV) and some particles accelerated near the footpoints (which did not 
have enough time to thermalise) can get to the photosphere (i.e., the lower boundary of the simulation domain). 

\subsection{Particle energy spectra}\label{tp-ensp}

There are two main factors affecting particle energies: parallel electric field and Coulomb scattering. The appearance of a 
strong electric field is a transient and local effect, while collisional deceleration of high-energy particles is 
always present. Hence, 
unlike in the collisionless models by \citet{gore11} and \citet{gobr12}, it is not possible 
to characterise the acceleration process by the final spectra in this type of model, as these spectra are inevitably 
thermal. We consider the energy spectra at three different stages 
of the reconnection process: just after the kink instability occurs, during the fastest energy release (i.e., 
when $d\mathcal{E}_m/dt$ is highest), and during the decay stage. The energy spectra for protons and electrons for low- and 
high-density cases are shown in Figs.~\ref{f-esplow}, \ref{f-esphigh}, \ref{f-psplow}, and \ref{f-psphigh}.

The electron energy distributions at the beginning of reconnection are similar to those obtained in simulations with 
no collisions  \citep{gobr11}: the spectra are combinations of a Maxwellian thermal distribution and nearly power-law
high-energy tail. However, in the low-density case the high-energy tail, surprisingly, appears 
to be softer than in previous studies; its power-law index is about $2.0-3.5$, which is observed in many flares. In 
the high-density case, the ``high-energy'' tail is harder, the spectral index is about $1.5-2.0$. At the later 
stages ($t\;=\;1100\; -\; 1500\;t_0$), the electric fields gradually 
decay and the collisions become dominant. This results in the hardening of the spectra around a few keV, and at some point,
a gap appears between the thermal part and high-energy part. This spectral hardening at lower energies (of few deka-keV) is
similar to that, which appears in thick-target models.

Comparing electron and proton energy spectra demonstrates the contrast in acceleration times: electron non-thermal 
spectra are formed within
$\sim 1-10 \;t_0$ after a kink, while it takes around $10-100 \, t_0$ to form a smooth high-energy tail for protons.
At the same time, protons lose their energies due to collisions faster than electrons with the same energy 
(since $d\mathcal{E}/dt \sim \sqrt{m/\mathcal{E}}$). As a result, it is more difficult in a collisional plasma to 
accelerate protons than electrons. 
The difference in proton and electron acceleration efficiencies can also be 
explained in terms of the electric field required to accelerate particles with energy $\mathcal{E}$ in presence of collisions
in plasma with density $n$. Indeed, particles are accelerated if $e E > 2 \pi e^4 \Lambda \frac m{m_e} n / \mathcal{E}$, 
or if the minimum field required to accelerate particles is 
\[
E_{crit} > 2 \pi e^3 \Lambda \frac m{m_e} n / \mathcal{E}.
\]
(For thermal electrons ($ \mathcal{E}=k_B T$, $m=m_e$), this becomes the standard expresion for Dreicer field.) 
Obviously, this $E_{crit}$ value for protons is higher by a factor of $m_p/m_e$ than $E_{crit}$
for electrons. Therefore, when the typical electric field in the system is higher than both $E_{crit}$(protons) and 
$E_{crit}$(electrons), the proton acceleration efficiency should be similar to that for electrons. However, when the typical 
electric field is lower than $E_{crit}$(protons) but higher than $E_{crit}$(electrons), the number of high-energy protons should be 
lower than that of electrons, which is the case in the Model B.

This difference between the low-density and high-density cases can also be seen in total energetics. Thus, the total energy carried by high-energy protons and electrons in the Model A (low-density case) is nearly equal and is about $6-8\%$ of the energy 
released during the reconnection, (although the maximum total non-thermal proton energy is reached slightly later then 
maximum total non-thermal electron energy). At the same time, the total energy carried by non-thermal electrons in the 
high-density case (Model B) reaches $\sim 4\%$ of the energy released during
reconnection, while protons carry less than $1\%$. These values, however, show the energies contained in non-thermal particles 
at a given moment and do not necessarily reflect the total energies {\it deposited} by non-thermal particles (or, in other 
words, the electromagnetic energy converted into thermal energy through non-thermal particles). This would depend on the value of the electric field compared to the local $E_{crit}$ value. Thus, particles accelerated by electric field $E \sim E_{crit}$ are 
almost instantly thermalised, which means that the number of non-thermal particles in the system at any time remain low, 
although a substantial amount of energy is converted from electric to thermal in this process.

Here, we define non-thermal as particles with energies $>1$~keV (the temperature of the initial distribution corresponds to 
around $100$~eV). The energy spectrum below this energy is not correct in the present model, since Eq.~\ref{gca-emslie} is 
only valid for non-thermal electrons with an energy that is much higher than $k_BT$ for the ambient plasma.

\begin{figure*}
\centerline{\includegraphics[width=0.6\textwidth,height=0.25\textheight]{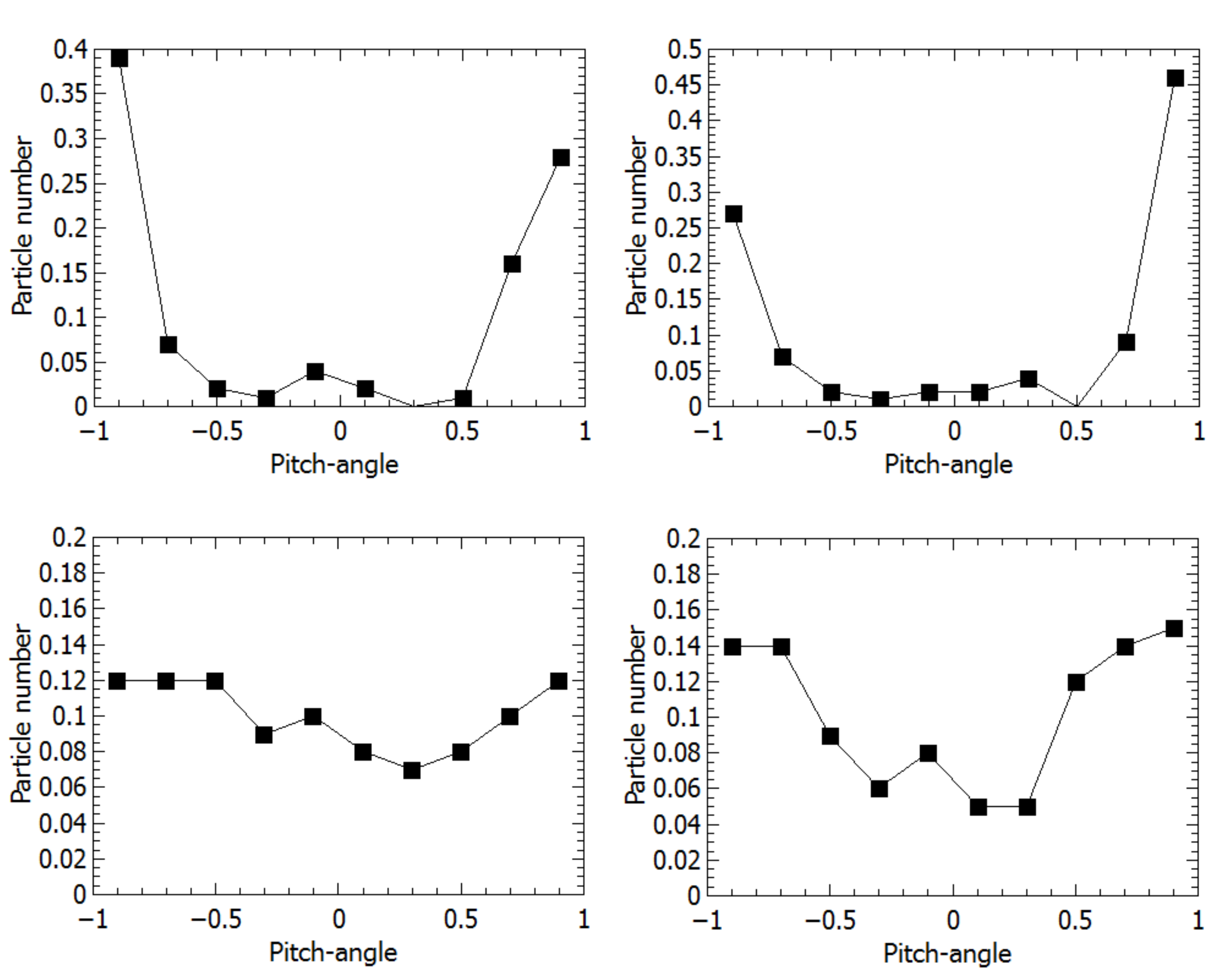}}
\caption{Normalised pitch-angle distributions of high-energy particles ($\mathcal{E} > 5$~keV) during fast reconnection 
stage ($t=1120 t_0$). Left panels (a and c) are for electrons; 
right panels (b and d) are for protons. Upper panels (a and b) correspond to the low-density case (Model A), while  
lower panels correspond to the high-density case (Model B).}
\label{f-pang}
\end{figure*}

\subsection{Particle pitch-angle distributions}\label{tp-ensp}

In collisionless models of particle acceleration by parallel electric fields, the pitch-angle distribution is trivial: 
since perpendicular velocities remain $\sim v_{th}$ and parallel velocitites increase drastically, all non-thermal particles 
are strongly collimated along magnetic field lines (i.e., $v_g/v_{||} << 1$), apart from particles being reflected by 
converging magnetic field. Collisions, however, result 
in effective scattering of supra-thermal particles with moderate energies at $\sim 1-10$~keV. Pitch-angle distributions are shown in 
Figure \ref{f-pang}.

It can be seen that particles in the low-density case (Fig.\ref{f-pang} a-b) are also well-collimated along 
the field lines, although the collimation is not as strong as in collisionless models. The majority of 
particles ($\sim 80\%$) have pitch-angles $|\alpha| > 0.75$.
In the high-density case, however, pitch-angle distributions are more isotropic: $\sim 40-50\%$ of particles have
pitch-angles of $|\alpha| <0.75$. 

The pitch-angle distributions also reveal the preferred directions of acceleration for diferent species. Thus, protons tend to have positive pitch-angles indicating that they move predominantly towards $y=+6.4L_0$ footpoint, while electrons tend to have negative
pitch-angles and, hence, move towards the opposite footpoint. This effect can also be seen in spatial distribution (see next section).
Coulomb collisions strongly affect particles in Model B, and, as the result, the proton-electron footpoint
asymmetry almost disappears in the high-density case. 

\begin{figure}
\centerline{\includegraphics[width=0.48\textwidth]{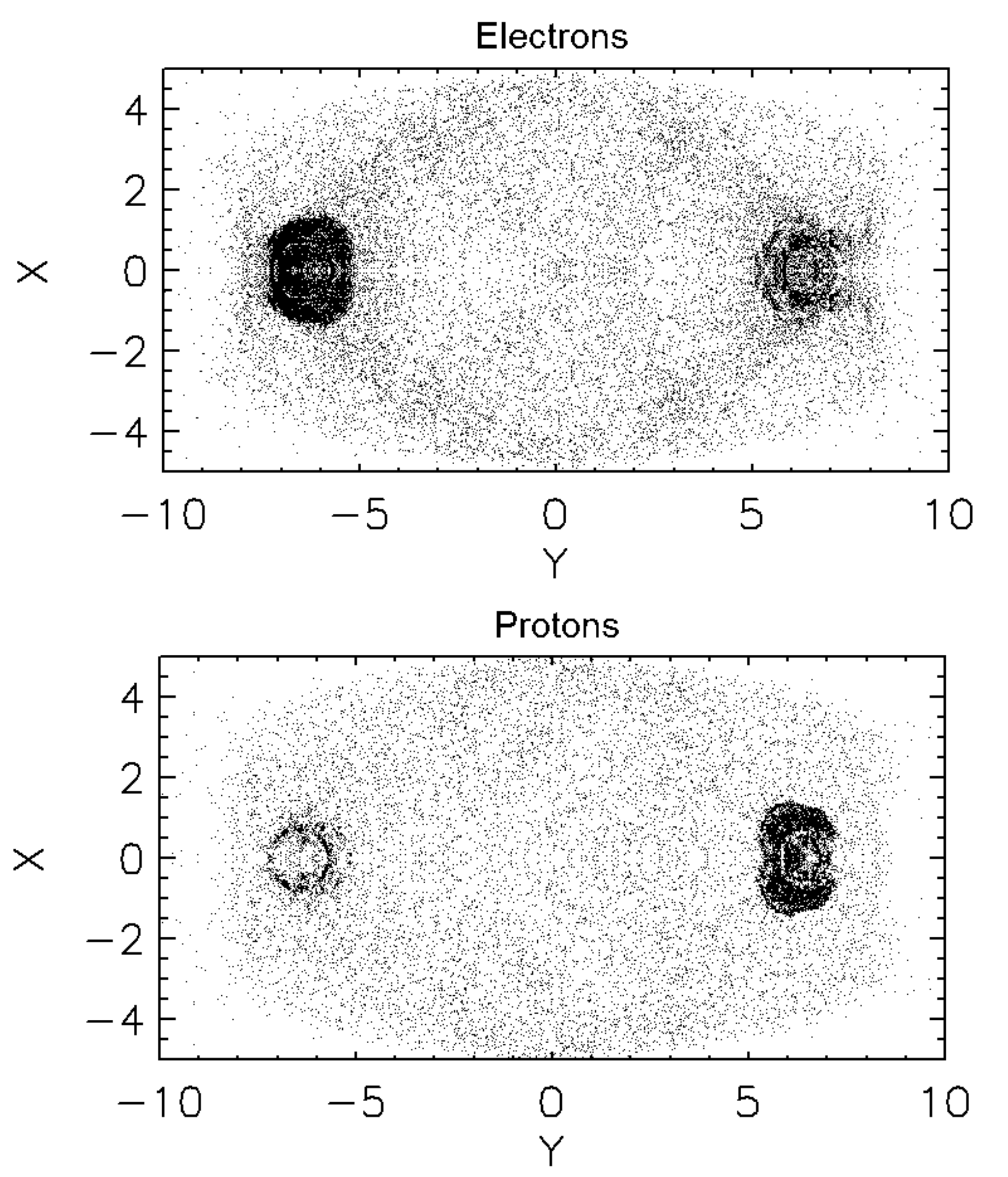}}
\caption{High-energy ($\mathcal{E} > 5$~keV) electron and proton distribution in $X-Y$ plane. Most of the particles are 
concentrated at low heights close to the footpoints.}
\label{f-pdistr}
\end{figure}

\begin{figure*}
\centerline{\includegraphics[width=0.65\textwidth]{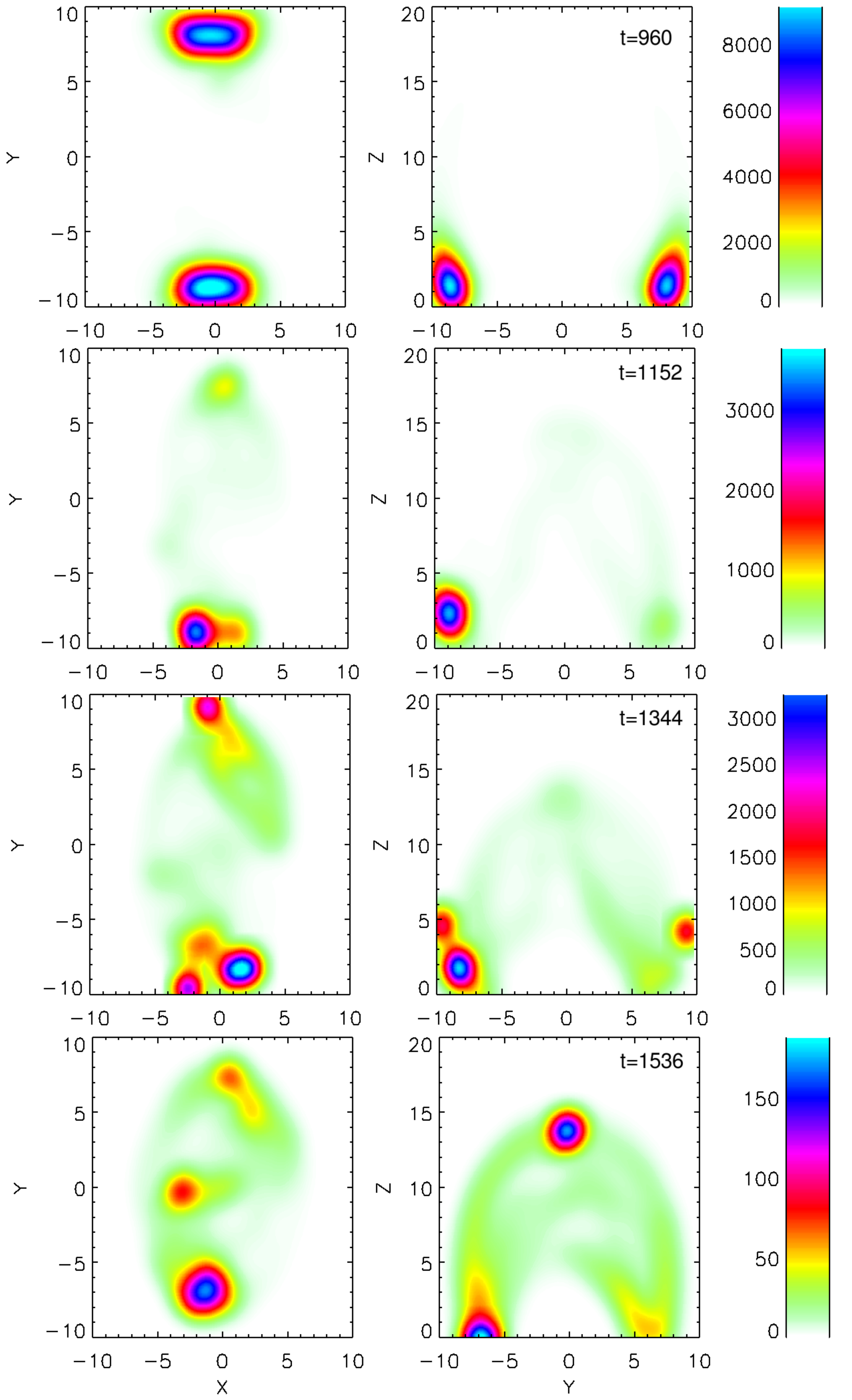}}
\caption{Synthetic hard X-ray emission at $10$~keV from the simulated flaring loop in low-density atmosphere.}
\label{f-xraylow}
\end{figure*}

\begin{figure*}
\centerline{\includegraphics[width=0.65\textwidth]{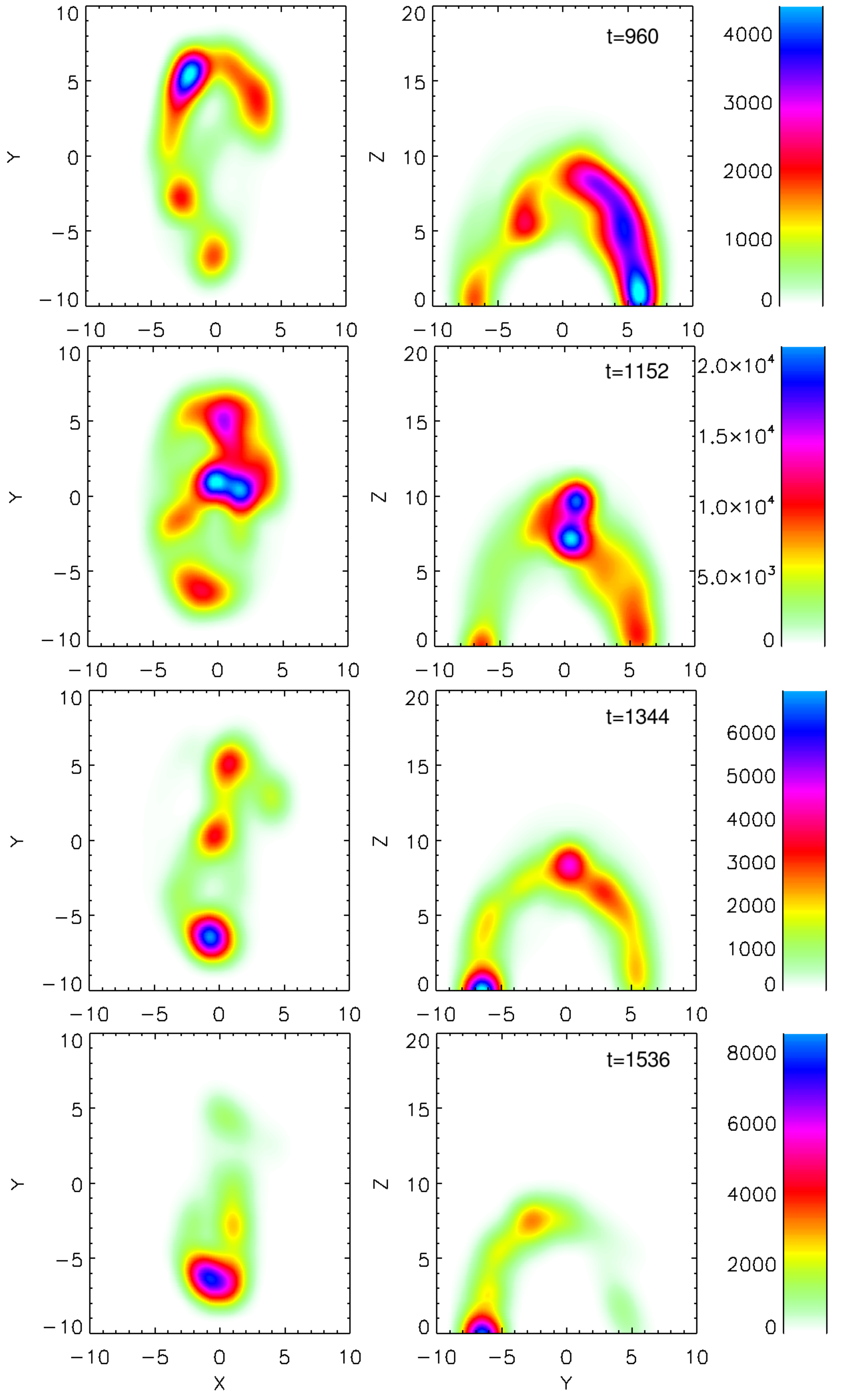}}
\caption{Synthetic hard X-ray emission at $10$~keV from the simulated flaring loop in high-density atmosphere.}
\label{f-xrayhigh}
\end{figure*}

\subsection{Particle spatial distribution and synthetic hard X-ray intensities}\label{tp-spat}

The helicity does not change its sign in the initial configuration, which means the current has preferential direction. This results 
in charge separation at the early stages of reconnection: electrons and protons are accelerated towards different footpoints, 
producing the electron-proton asymmetry of footpoints that can be seen in Figure~\ref{f-pdistr}. As 
the current structure becomes filamentary with time, the acceleration process becomes more chaotic and footpoint asymmetry
decreases. At the end of reconnection, only a slight assymetry can be seen. This is consistent with observations 
showing time delay in appearance of a second footpoint source.

Figure~\ref{f-pdistr} also shows that the footpoint area (which represents the cross-section of the volume occupied by high-energy 
particles just above the photosphere) increases with time. During the reconnection, the radius of footpoints increases from $\sim 0.5$~Mm to nearly $1.2$~Mm. This effect is mainly due to the reconnection between the field lines of the 
twisted fluxtube and ambient field lines \citep[see discussion in][]{gobr11}.

Based on the electron distributions derived in Models A and B, we calculate 
intensities of the bremsstrahlung hard X-ray emission from the flaring loop. We use the simplified form 
of the bremsstrahlung cross-section \citep[see e.g.][]{kone02}:

\begin{equation}
I(\epsilon)=const \int \limits_\epsilon^\infty \int \limits_L N(E, l) \frac {n(l)}{\epsilon \sqrt{E}}dE dl.
\end{equation}

The synthetic intensity maps are shown in Figures~\ref{f-xraylow} and \ref{f-xrayhigh} for
four different moments. In order to compare them with observational data, they are smoothed by a Gaussian profile with the half-width of $\sim 1.5$~Mm (comparable to the RHESSI spatial resolution). The main difference between the low-density case (Model A) and 
high-density case (Model B) is that 
the latter shows noticeable extended emission from the loop, while the low-density case has noticeable emission only from the 
footpoints. This can be easily explained by comparing the electron mean-free-path in these two models: in the low-density model, 
the mean free path for the energies $\sim 10$~keV is substantially longer than the loop length ($\sim 20$~Mm), while it is only 
$\sim 1$~Mm in the high-density case.

\section{Summary}\label{concl}

In the present work, we use a combination of MHD and test-particle methods to study energy release and high-energy particle
motion in a reconnecting twisted coronal loop. The approach is generally similar to that previously used in
2D \citep{gore11} and 3D models \citep{gobr11,gobr12}: time-dependent electric and magnetic fields and density distributions are used 
as an input for guiding-centre test-particle calculations of proton and electron trajectories. The main benefits from the present 
model are that
\begin{itemize}
\item we use a realistic loop-like structure \citep[similarly to][]{klie10}, and the twist is created by slow footpoint rotation;
\item we take into account strong stratification in the lower atmosphere and consider different densities in the corona;
\item we take into account the effect of Coulomb collisions on test-particle motion. 
\end{itemize}
Therefore, this study attempts to combine energy release during reconnection with particle acceleration 
and transport within a single model. Obviously, the main problem is that the feedback of energetic particle motion 
on the electric and magnetic field evolution is 
ignored. However, this should not significantly affect evolution of the system in the cases studied here, since the acceleration efficiency is low: high energy particles carry less than $\sim 10\%$ and hence, the plasma is almost ``thermal'' \citep[see also][]{gore12}. 

Magnetic reconnection in the present models is similar to that observed in cylindrical fluxtubes 
with analytical magnetic field twist profiles 
\citep[e.g.][]{hooe09,gobr11}. Thus, a strong, helically shaped current is formed after the kink occurs; the 
current stucture gradually becomes filamentary as the reconnection proceeds. The connectivity changes in two different ways: first, 
the magnetic twist is reduced and second, as shown by \citet{gobr11}, the twisted loop field lines reconnect with ambient field lines,
resulting in an effective increase in the loop cross-section. However, there are several features in the present model, which are not observed in models with cylindrical fluxtubes. First, a current torus, almost perpendicular to the loop, is formed near the loop-top just after the kink. Second, although the magnetic energy dissipates in three extended regions (near the footpoints and the 
loop-top), the field connectivity changes mainly due to magnetic reconnection just below the magnetic torus, which is formed near 
the loop-top after the kink. Finally, dissipation of the azimutal field leads to
a substantial shortening of the loop field lines.
The density of the plasma in the corona does not seem to have a noticeable effect on reconnection, as expected, since the plasma
beta is low (around $10^{-1}$) in both cases.

Particles can be accelerated almost anywhere within the reconnecting loop, although they are predominantly energised close to the
loop-top, where the ratio of electric field to Dreicer field is the highest. The energy spectra are combinations of Maxwellian thermal 
distributions at low energies (below keV) and are nearly power-low high-energy tails. The spectral indices for electrons in the
$10-100$~keV range are between $1.5$ and $3.0$, which means the spectra are slightly softer than in similar collisionless models. The electron spectra are harder in the case with dense coronal plasma ($10^{11}$ cm$^{-3}$) than in the case with normal coronal density 
($10^9$ cm$^{-3}$). The acceleration efficiency strongly depends on the ambient plasma density. This is not surprising, since the 
particle acceleration generally depends on the ratio of electric field to the so-called critical electric field 
$E_{crit} \sim m n/\mathcal{E}$ (i.e. the
minimum electric field required to produce runaway particles; it is equal to the Dreicer field in case of thermal electrons) (see 
Sect.~\ref{tp-ensp}). In both cases 
(low- and high-density corona), the electric field is higher than the corresponding
$E_{crit}$ value for electrons. However, in the high-density case the electric field in the system is comparable 
to the corresponding $E_{crit}$ value for protons, and there is almost no proton acceleration. This results in an interesting picture: in the
low-density medium, the non-thermal particle energy is nearly equally split between protons and electrons, while in the high-density 
medium, protons remain nearly thermal amid a substantial number of non-thermal electrons. This can potentially explain why some flares
effectively accelerate both electrons and ions, and other flares produce substantial number of electrons,
while not showing any noticeable ion emission. Many phenomenological models suggest that this happens because, even though there are plenty of ions, their energy is insufficient to produce nuclear de-excitation lines needed for detection. However, it might be
the case that ions in electron-only flares do not get accelerated due to high plasma density in the acceleration region.

High-energy particles are rather uniformly distributed along the flaring loop. There is a noticeable increase in the particle number 
towards footpoints, which can be easily explained by lower average parallel velocities due to magnetic field convergence. There
is a visible proton-electron asymmetry between footpoints at the early stages of acceleration, because the initial configuration has
a non-zero net current and hence, the electric field has a preferential direction along the fluxtube before it gets filamentary. Although this proton-electron asymmetry would result in the development of an oppositely-directed large-scale electric field along 
the reconnecting loop (the so-called return current electric field), this effect can still exist, as the non-thermal 
proton-electron current can be compensated by an ambient thermal electron current, provided the plasma is highly ionised and 
the number of high-energy particles is low (which is the case here) \citep[see][]{zhgo04,zhgo05}.
In the high-density case, this effect is less noticeable owing to pitch-angle scattering due to Coulomb collisions.

In general, the structure of the synthetic hard X-ray sources is similar to that observed with RHESSI  \cite[e.g.][]{lin11}. 
There are high-intensity
footpoint sources, which are flat (their height is only about $0.5-1.0$~Mm), and 
weaker extended emission from the whole loop. The
ratio of the footpoint and extended source intensities depends on the plasma density in the loop. In the case of normal coronal 
density, the emission comes mostly from the foopoints, while the intensity of the extended source in the high-density case is
comparable to the footpoint intensities. The size of the footpoint sources increases with time (due to expansion of the twisted loop
as discussed in Sect.~\ref{tp-spat}; however, this effect is less prominent in hard X-ray due to the assumed temporal and spatial 
resolution of the instrument. 

Similar to \citet{gobr12}, we assume that the appearance of a strong parallel electric field due 
to anomalous resistivity in fragmented current structures is the main mechanism that accelerates particles. Other mechanisms, such as 
acceleration due to magnetic field contraction and large scale MHD waves, are included within our model and can have an effect, although they are found to be negligible
compared to the parallel electric field (for the assumed value of anomalous resistivity and the adopted length scale). 

Our work suggests that large-scale MHD turbulence develops during magnetic reconnection and results in fragmentation of initially solid current \citep[see also][]{broe08,hooe09}. This phenomenon affects particle behaviour: proton and electron trajectories become more stochastic, as they are accelerated through repeated encounters with current fragments (see Sect.~\ref{tp-traj} and \citet{gobr11,gobr12}, see also \citet{care12} for review). 
Another important factor, which can affect particle acceleration and transport is MHD turbulence at smaller spatial scales 
\citep[e.g.][]{mile96,lare96,chla06,cavl09,biko10,biae10}. Turbulence results in field line
stochasticity and fast reconnection in multiple small-scale current sheets \citep{lavi99,eyie11}. {Hence, the presence of small-scale turbulence, in principle, could drastically change characteristics of the electric field in the model and lead to a different picture of particle acceleration. The implications of turbulence for particle acceleration are summarized in \citet{laze12}.

Furthermore, turbulence can make substantial contribution to particle scattering during their transport. Scattering due to Coulomb collisions considered in the present model is expected to be dominant in the dense chromosphere and transition region. At 
the same time, small-scale MHD turbulence can be important for particle scattering in the low-density corona, along with various 
types of self-induced kinetic waves. Thus, high-energy ions have been shown to generate kinetic Alfven waves \citep[e.g.][]{voit98}, which, in turn, result in noticeable heating in the corona \citep{gore05} (while collisional scattering results only in effective 
heating of dense transition region and chromosphere).

However, it is impossible to include these microscopic effects in our large-scale model. The macroscopic description of these 
microscopic effects is required for them to be included in large-scale transport models, which necessitates more studies 
into this issue.

The scenario developed in the present paper is a good candidate for interepreting small non-eruptive explosive events occuring in single loop configurations. A benefit of this model is reduced energy losses during particle transport due to more uniform (rather than localized at the 
loop-top) acceleration \citep[see discussion in][]{gore12}. The model, however, needs to be further improved; in particular, one needs to consider the possibility of continuous helicity injection (or footpoint twisting).

\begin{acknowledgements}
This work is funded by Science and Technology Facilities Council (UK). 
\end{acknowledgements}


\begin{thebibliography}{00}
\bibitem[\protect\citeauthoryear{Amari et al.}{2000}]{amae00} Amari, T., Luciani, J.F., Mikic, Z. \& Linker, J., 2000, ApJ, 529, L49
\bibitem[\protect\citeauthoryear{Arber et al.}{2001}]{arbe01} Arber, T.G., Longbottom, A. W., Gerrard, C. L. \& Milne, A. M. 
2001, J.Comp.Phys, 171, 151
\bibitem[\protect\citeauthoryear{Archontis et al.}{2010}]{arce10} Archontis,  V., Tsinganos, K. \& Gontikakis, C., 2010, A\&A, 512, L2
\bibitem[\protect\citeauthoryear{Aschwanden et al.}{2009}]{asce09} Aschwanden, M.J., Wuelser, J.P., Nitta, N.V. \& %
Lemen, J.R. 2009, Solar Phys., 256, 3
\bibitem[\protect\citeauthoryear{Bareford et al.}{2013}]{bare13} Bareford, M. R., Hood, A.W. \& Browning, P.K., 2013, A\&A, 550, 40
\bibitem[\protect\citeauthoryear{Barta et al.}{2011}]{bare11} Barta, M., Buechner, J., Karlicky, M. \& Skala, J. 2011, ApJ%
737, 24
\bibitem[\protect\citeauthoryear{Bian and Browning}{2008}]{bibr08} Bian, N.H. \& Browning, P.K., 2008, ApJ, 687, L111
\bibitem[\protect\citeauthoryear{Bian and Kontar}{2010}]{biko10} Bian, N.H. \& Kontar, E.P., 2010, Phys. Plasmas, 17, 062308
\bibitem[\protect\citeauthoryear{Bian  et al.}{2010}]{biae10} Bian, N.H., Kontar, E.P. \& Brown, J.C., 2010, A\&A, 519, A114
\bibitem[\protect\citeauthoryear{Brown et al.}{2003}]{broe03} Brown, D.S., Nightingale, R.W., Alexander, D., Schrijver, C.J. 
Metcalf, T.R. Shine, R.A., Title, A.M. \& Wolfson, C.J., 2003, Solar Phys., 216, 79
\bibitem[\protect\citeauthoryear{Brown}{1976}]{brow76} Brown, J.C., 1976, Phil. Trans. Roy. Soc. Lond. A, 281, 473
\bibitem[\protect\citeauthoryear{Brown et al.}{2009}]{broe09} Brown, J.C., Turkmani, R., Kontar, E.P., MacKinnon, A.L. \& Vlahos, L., 
2009, A\&A, 508, 993
\bibitem[\protect\citeauthoryear{Browning and Van der Linden}{2003}]{brva03} Browning, P.K. \& van~der~Linden, R.A.M., 2003, A\&A, 
400, 355
\bibitem[\protect\citeauthoryear{Browning et al.}{2008}]{broe08} Browning, P.K., Gerrard, C., Hood, A.W., Kevis, R. \& 
van~der~Linden, R.A.M., 2008, A\&A, 485, 837
\bibitem[\protect\citeauthoryear{Cargill et al.}{2012}]{care12} Cargill, P.J., Vlahos, L., Baumann, G., Drake, J.F. \& Nordlund, A., %
2012, Space Sci. Rev., 173, 223.
\bibitem[\protect\citeauthoryear{Cargill and Vlahos}{2009}]{cavl09} Cargill, P.J. \& Vlahos, L., 2009, Lect. Not. Phys., v.778.
\bibitem[\protect\citeauthoryear{Cho and Lazarian}{2006}]{chla06} Cho, J. \& Lazarian, A. 2006, ApJ, 638, 811
\bibitem[\protect\citeauthoryear{Emslie}{1978}]{emsl78} Emslie, A.G., 1978, ApJ, 224, 241
\bibitem[\protect\citeauthoryear{Eyink et al.}{2011}]{eyie11} Eyink, G., Lazarian, A. \& Vishniak, E.T. 2011, ApJ, 743, 51
\bibitem[\protect\citeauthoryear{Fan}{2010}]{fan10} Fan, Y., 2010, ApJ, 719, 728
\bibitem[\protect\citeauthoryear{Gibson and Fan}{2006}]{gibe06} Gibson, S.E. \& Fan, Y., 2006, ApJ, 637, L65
\bibitem[\protect\citeauthoryear{Gordovskyy and Browning}{2011}]{gobr11} Gordovskyy, M. \& Browning, P.K. 2011, ApJ, 729, 101
\bibitem[\protect\citeauthoryear{Gordovskyy and Browning}{2012}]{gobr12} Gordovskyy, M. \& Browning, P.K. 2012, Solar Phys., 
277, 299
\bibitem[\protect\citeauthoryear{Gordovskyy et al.}{2012}]{gore12} Gordovskyy, M., Browning, P.K., Kontar, E.P. \&%
Bian, N.H. 2012, Solar Phys., accepted
\bibitem[\protect\citeauthoryear{Gordovskyy et al.}{2010}]{gore10} Gordovskyy, M., Browning, P.K. \& Vekstein G.E. 2010, A\&A, 519, A21
\bibitem[\protect\citeauthoryear{Gordovskyy et al.}{2011}]{gore11} Gordovskyy, M., Browning, P.K. \& Vekstein G.E. 2011, ApJ, 720, 1603
\bibitem[\protect\citeauthoryear{Gordovskyy et al.}{2005}]{gore05} Gordovskyy, M., Zharkova, V.V., Voitenko, Yu.M. \& Goossens, M. 2005, Adv. Space Res., 35, 1743
\bibitem[\protect\citeauthoryear{Hood et al.}{2009}]{hooe09} Hood, A.W., Browning, P.K., van der Linden, R.A.M. 2009, A\&A, 
506, 913
\bibitem[\protect\citeauthoryear{Kliem et al.}{2000}]{klie00} Kliem, B., Karlicky, M. \& Benz, A.O. 2000, A\&A, 360, 715
\bibitem[\protect\citeauthoryear{Kliem et al.}{2010}]{klie10} Kliem, B., Linton, M.G., Toeroek, T., Karlicky, M. 2010, %
Solar Phys., 266, 91 
\bibitem[\protect\citeauthoryear{Kontar et al.}{2002}]{kone02} Kontar, E.P., Brown, J.C. \& McArthur, G.K, 2002, Solar Phys., 210, 419
\bibitem[\protect\citeauthoryear{Kontar et al.}{2011}]{kone11} Kontar, E.P., Hannah, I.G. \& Bian, N.H., 2011, ApJ, 730, L22 
\bibitem[\protect\citeauthoryear{Kowal et al.}{2012}]{kowe12} Kowal, G., de Gouveia Del Pino, E.M. \& Lazarian, A. 2012, Phys.Rev.Lett., 19, 297
\bibitem[\protect\citeauthoryear{Larosa et al.}{1996}]{lare96} Larosa, T.N., Moore, R.L. \& Miller, J.A. 1996, ApJ, 467, 454L
\bibitem[\protect\citeauthoryear{Lazarian and Vishniak}{1999}]{lavi99} Lazarian, A. \& Vishniak, E.T. 1999, ApJ, 517, 70
\bibitem[\protect\citeauthoryear{Lazarian et al.}{2012}]{laze12} Lazarian, A., Vlahos, L., Kowal, G., Yan, H., Beresnyak, A. \& de Gouveia Dal Pino, E. M., 2012, Space Sci. Rev., 173, 557
\bibitem[\protect\citeauthoryear{Lin}{2011}]{lin11} Lin, R.P., 2011, Space Sci. Rev., 159, 421
\bibitem[\protect\citeauthoryear{Malanushenko et al.}{2011}]{male11} Malanushenko, A., Yusuf, M.H. \& Longcope, D.W., 2011,
ApJ, 736, 97 
\bibitem[\protect\citeauthoryear{Miller et al.}{1996}]{mile96} Miller, J.A., Larosa, T.N. \& Moore, R.L. 1996, ApJ, 461, 445
\bibitem[\protect\citeauthoryear{Northrop}{1963}]{nort63} Northrop, T. 1963, The Adiabatic Motion of Charged Particles 
(Interscience, New York)
\bibitem[\protect\citeauthoryear{Raouafi}{2009}]{raou09} Raouafi N.E. 2009, ApJ, 691, L128
\bibitem[\protect\citeauthoryear{Regnier and Amari}{2004}]{ream04} Regnier, S. \& Amari, T. 2004, A\&A, 425, 345
\bibitem[\protect\citeauthoryear{Srivastava et al.}{2010}]{srie10} 	Srivastava, A.K., Zaqarashvili, T.V., Kumar, P. \& 
Khodachenko, M.L. 2010, ApJ, 715, 292
\bibitem[\protect\citeauthoryear{Toeroek and Kliem}{2005}]{tokl05} Toeroek, B. \& Kliem, T., ApJ, 630, L97
\bibitem[\protect\citeauthoryear{Tripathi et al.}{2008}]{trie08} Tripathi, D., Mason, H.E., Young, P.R. \& Del Zanna, G. 2008, 
A\&A, 481, L53
\bibitem[\protect\citeauthoryear{Ugai}{1992}]{ugai92} Ugai, M. 1992, Phys.Plasm.B, 4, 2953 
\bibitem[\protect\citeauthoryear{Uzdensky}{2003}]{uzde03} Uzdensky, D. 2003, ApJ, 587, 450
\bibitem[\protect\citeauthoryear{Vernazza et al.}{1981}]{vere81} Vernazza, J.E., Avrett, E.H. \& Loeser, R. 1981, ApJ Suppl., 
45, 635
\bibitem[\protect\citeauthoryear{Voitenko}{1998}]{voit98} {Voitenko, Yu.M. 1998, Solar Phys., 182, 411}
\bibitem[\protect\citeauthoryear{Zharkova and Gordovskyy}{2004}]{zhgo04} Zharkova, V.V. \& Gordovskyy, M. 2004, ApJ, 604, 884
\bibitem[\protect\citeauthoryear{Zharkova and Gordovskyy}{2005}]{zhgo05} Zharkova, V.V. \& Gordovskyy, M. 2005, A\&A, 432, 1033
\end{thebibliography}
\end{document}